\documentclass[useAMS,usenatbib,fleqn]{mn2e}

\pdfoutput=1
\usepackage{amssymb}
\usepackage{amsmath}
\usepackage[english]{babel}
\usepackage{graphicx}
\title[Mass and spin coevolution during the alignment of a BH]{Mass and spin coevolution during the alignment of a black hole in a warped accretion disc}
\author[A. Perego et al.]{A. Perego$^{1}$\thanks{E-mail:
albino.perego@unibas.ch},  M. Dotti$^{2}$, M. Colpi$^{3}$, M. Volonteri$^{2}$\\
$^{1}$Department of Physics, University of Basel, Klingerbergstr. 82, 4056 Basel, Switzerland\\
$^{2}$Department of Astronomy, University of Michigan, Ann Arbor, MI 48109, USA\\
$^{3}$Dipartimento di Fisica, Universit\`a degli Studi di Milano-Bicocca, Piazza Della Scienza 3, 20126 Milano, Italy}

\begin{document}

\def\msun{\rm M_{\odot}}
\def\bh{{\rm BH}}
\def\bp{{\rm BP}}
\def\al{{\rm al}}
\def\w{{\rm warp}}
\def\outo{{{\rm out},0}}
\def\out{{{\rm out}}}
\def\discout{{\rm disc,out}}
\def\ed{{\rm Edd}}
\def\bfj{\bf{J}}
\def\acc{{\rm acc}}
\def\mbh{M_{\rm BH}}
\def\mbho{M_{\rm BH,0}}
\def\ISO{{\rm ISO}}
\newcommand{\be}{\begin{equation}}
\newcommand{\ee}{\end{equation}}

\date{Accepted 2009 July 20. Received 2009 July 17. In original form 2009 March 16}

\pagerange{\pageref{firstpage}--\pageref{lastpage}} \pubyear{2009}

\maketitle

\label{firstpage}

\begin{abstract}
In this paper, we explore the gravitomagnetic interaction of a black hole (BH) with a 
misaligned accretion disc to
study BH spin {\it precession} and {\it alignment} jointly with BH mass $M_\bh$ and  spin 
parameter $a$ evolution, under the assumption that the disc is continually fed, in its outer region, by matter with angular momentum fixed
on a given direction ${\hat \bfj}_{\rm disc,out}.$
We develop an iterative scheme based on the adiabatic approximation to
study the BH-disc coevolution: 
in this approach, the accretion disc transits through a sequence of quasi-steady warped states (Bardeen-Petterson effect) and interacts  
with the BH until the spin $\bfj_\bh$ aligns 
with  ${\hat \bfj}_{\rm disc,out}.$
For a BH aligning with a co-rotating disc, the fractional increase in mass is typically less than a few percent, while
the spin modulus can increase up to a few tens of percent. 
The alignment timescale $t_\al\propto {a}^{5/7}{\dot M}^{-32/35}$ is 
of $\sim 10^{5}-10^{6}$ yr for a maximally rotating BH accreting at the Eddington rate.  
BH-disc alignment from an initially counter-rotating disc
tends to be more efficient compared to the specular co-rotating case 
due to the asymmetry seeded in the Kerr metric: counter-rotating matter 
carries a  larger and opposite angular momentum when 
crossing  the innermost stable orbit, so that the spin modulus decreases 
faster and so the relative inclination angle. 

\end{abstract}


\section{Introduction}

Astrophysical black holes (BHs) are Kerr black holes fully characterized by 
their mass $M_\bh$ and spin $\mathbf{J}_\bh$, customarily expressed 
in terms of the dimensionless spin parameter $a$  ($\leq 1$), and unit 
vector ${\hat {\bfj}}_\bh$:
\begin{equation} \label{eqn:jbh definition}
  \mathbf{J}_\bh=a\frac{G M_\bh^2 }{c}\hat{\mathbf{J}}_{\bh}.
\end{equation}
The spin and mass of BHs residing 
in galaxy nuclei do not remain constant, close to their birth values, 
but change sizeably through cosmic time, in response to major accretion events. 
In current cosmological scenarios for
the evolution of galaxies, repeated interactions 
among  gas-rich halos play a key role not only in shaping galaxies, but also 
in triggering quasar activity \citep{WhiteRees1978, DiMatteoSpringelHernquist2005}.
Massive gaseous nuclear discs that form in the aftermath of major galaxy 
mergers \citep{MihosHernquist1996, Mayeretal2007} may provide
enough fuel 
to feed, on sub-parsec scales,
the BH through a Keplerian accretion disc \citep{Dottietal2007, Dottietal2009}.
If these episodes repeat recursively and/or at random phases 
\citep{KingPringle2006} the BH spin 
$\mathbf{J}_\bh$ is expected, initially, to be misaligned 
relative to the direction of the angular momentum of the disc
${\hat \bfj}_{\rm disc,out}$ at its unperturbed, outer edge $R_{\rm out}$.
In this configuration, the gas elements inside the disc undergo Lense-Thirring precession \citep[see, e.g. ][]{Wilkins1972}. 
In the fluid, the action of viscosity onto the differentially precessing 
disc ensures that the inner portion of 
the accretion disc aligns (or anti-aligns) its orbital angular momentum with 
the BH spin $\bfj_\bh$, out to a transition radius $R_\w$  beyond which 
the disc remains aligned to the outer disc, 
as first shown by \citet{BardeenPetterson1975}\citep[see also][]{ArmitageNatarajan1999,NelsonPapaloizou2000,FragileAnninos2005,Fragileetal2007}.  
Warping of the inner disc at distance $R$ from the BH is communicated through 
the fluid elements on a timescale $t_\bp(R)$ related to the vertical 
shear viscosity of the accretion disc. 
Therefore, 
the inner regions of the disc align (or counter-align if
the disc is counter-rotating) with the BH spin
on the scale $t_\bp(R_\w)$ when the viscous time 
for vertical propagation of disturbances equals the Lense-Thirring precession 
time.
On a longer timescale, the joint evolution of BH+disc system restores full 
axisymmetry, with the BH spin direction aligned relative to the
total angular momentum of the composite system  \citep{Rees1978, ThornePriceMacDonald1986, Kingetal2005}. 
The change in $\bfj_\bh$ is a consequence of angular momentum conservation:
since the BH acts on the disc with a torque that warps the disc, then an equal and opposite gravito-magnetic torque acts on the BH that modifies its direction {\it only}.  

BH spin alignment has been studied in two main contexts. 
In the first, explored by \citet{Kingetal2005} and \citet{LodatoPringle2006}, 
the focus is on an {\it closed} system where the 
accretion disc has a finite mass and radial extent. 
Here, the total angular momentum
$\bfj_{\rm tot}=\bfj_\bh+\bfj_{\rm disc}$ is well defined vector, and the BH eventually
aligns it spin vector to the direction of $\bfj_{\rm tot}$.
In the second, explored by \citet{ScheuerFeiler1996}, \citet{NatarajanArmitage1999} and \citet{MartinPringleTout2007},
the focus is on an {\it open} system, where the accretion disc has infinite
extension and it is continually fed at its outer edge by 
matter whose angular momentum has constant direction 
$\hat{\bfj}_{\rm disc,out}$. 
In this second case, the BH aligns its spin to the outer disc direction $\hat{\bfj}_\discout$ on a timescale  
$t_\al$ that exceeds $t_\bp$ by a few orders of magnitude \citep{ScheuerFeiler1996,MartinPringleTout2007}.  

In this paper, we progress on the study of BH alignment including
the contemporary
change in mass and spin modulus due to accretion of matter, neglected in previous works.
During BH precession and alignment, matter flows inward and accretes
carrying the energy and the specific angular momentum of the innermost 
stable circular orbit (ISO).
This study thus provides estimates of the fractional increase 
of mass $\Delta M_\bh/M_{\bh,0}$ and spin $\Delta {a}/{a}_0$ during 
BH alignment (the subscript $0$ refers to initial conditions), 
together with a sensible expression for the alignment time $t_\al$. 
In our context, we assume a continuous and coherent feeding of the accretion 
disc around the BH, at least for a time as long as the alignment timescale 
$t_{\rm al}$. Thus, we consider an open system and we fix the orbital 
angular momentum direction $\hat{\bfj}_\discout$ at the outer edge of the disc.

In Section 2, we introduce key parameters and highlight our model
assumptions.  
Section 3 surveys properties of steady state
warped discs and key scales associated to the Bardeen-Petterson 
effect; disc models with constant and power-law viscosity profile are
explored for completeness. In Section 4, we describe the equations for the 
BH mass and spin evolution, and introduce 
the adiabatic approximation to solve these equations.
In the same section we also revisit the expression for the BH alignment time.
Results are illustrated in Section 5; there we explore also the
tendency to alignment in initially counter-rotating warped discs. 
Section 6 contains the discussion of the results and our conclusions.

\section{Initial assumptions and main parameters}
\label{sec:initial assumptions}

We consider a BH 
with spin ${\bfj_\bh}$, surrounded by a geometrically thin, standard 
Shakura-Sunyaev $\alpha$-disc \citep[e.g.][]{ShakuraSunyaev1973,
FrankKingRaine2002}. The $\alpha$-disc is initially misaligned
relative to $\bfj_\bh$, i.e. the  angular momentum unit vector of the disc  at the 
outer edge is $\hat{\bf{J}}_{\discout}\neq {\hat{\bfj}}_\bh$; the 
relative inclination angle between the two unit vectors is $\theta_{\out}$.

Following \citet{Pringle1992}, we assume that the accretion disc has a high 
viscosity ($\alpha > H/R$, where $H$ is the disc vertical scale height) so that 
perturbations propagate diffusively. 
We introduce two viscosity parameters, 
$\nu_1$ and $\nu_2$: $\nu_1$ is the standard radial shear viscosity 
while $\nu_2$ is the vertical share viscosity associated to the diffusion 
of vertical warps through the disc, due to Lense-Thirring precession. 
For $\nu_1$ we adopt the $\alpha$ prescription
\begin{equation} \label{eqn:alpha-prescription}
     \nu_1=\alpha H c_{\rm s}
\end{equation}
where $c_{\rm s}$ is the sound speed inside the accretion disc.
 It is still poorly understood which is the relation between the radial
and the vertical viscosity; in particular, if $\nu_1 \sim \nu_2$ or 
$\nu_1 \ll \nu_2$. In order to simplify our discussion, we refer 
to the recent analysis of \citet{LodatoPringle2007}, and for $\nu_2$ we take: 
\begin{equation} \label{eqn:nu1-nu2 relation}
  \frac{\nu_2}{\nu_1}=\frac{f_{\nu_2}}{2\alpha^2}
\end{equation}
where $f_{\nu_2}$ (given in Table 1) is a coefficient determined in numerical simulations that 
accounts for non-linear effects.
 
The disc model is defined after specifying five free parameters (subscript 0
will be introduced to indicate initial values when mass and
spin evolution is considered):

\noindent
(1)  The BH mass, $\mbh$; we explore
a mass range between $10^5 \msun < \mbh < 10^7 \msun$. 
For the BH mass we introduce 
the dimensionless parameter $M_6$ as $ \mbh = M_6 \times 10^6 \msun.$

\noindent
(2) The spin modulus, in terms of the dimensionless spin 
parameter $a$,
which varies between  $0 < a \leq 0.95$. We do not use the theoretical 
limit $a=1$ because, if accretion is driven by magneto-rotational 
instabilities in a relativistic MHD disc, the final equilibrium spin due 
to continuous accretion is 
$a \approx 0.95$ \citep{GammieShapiroMcKinney2004};

\noindent
(3) The relative inclination angle $\theta_\out,$ 
between the spin versor $\hat{\bfj}_\bh$ and the orbital angular momentum 
versor at the external edge of the accretion disc, $\hat{\bfj}_\discout$. 
This angle varies isotropically from 0 to $\pi$. In the following, however, 
we will confine this interval to ($0,\sim \pi/6$) in order to satisfy the 
used approximations.

\noindent 
(4) The viscosity parameter $\alpha$ which is assumed to vary
between $10^{-2}\lesssim \alpha \lesssim 10^{-1}$ to bracket 
uncertainties \citep{KingPringleLivio2007}.  
For our purposes we selected values of $\alpha$ 
according to \citet{LodatoPringle2007}, as in Table \ref{tab:viscosita' lodato}. In this study, $\alpha$ is considered as a constant inside
the disc. 
 \begin{table}
   \begin{center}
     \begin{tabular}{|c|c|}
       \hline
      $\alpha$ & $f_{\nu_2}$\\
      \hline
      0.18 & 1.00 \\
      0.15 & 0.85 \\
      0.09 & 0.60 \\
      0.05 & 0.38 \\
      \hline
    \end{tabular}
    \caption{Table of the coefficients $\alpha$ and $f_{\nu_2}$.}
    \label{tab:viscosita' lodato}
  \end{center}
\end{table}

\noindent
(5) The accretion rate onto the BH, $\dot{M},$ is expressed 
in terms of the Eddington ratio $f_\ed=L/ L_\ed$ and of the accretion efficiency $\eta$
(where $L_\ed$ is the Eddington luminosity):
${\dot M}=f_\ed L_\ed/(\eta c^2)$.
We consider values of $f_\ed$ in the interval 
$ 10^{-4} < f_\ed < 1$  and compute $\eta$ as a function of the BH spin modulus.

If the disc, warped in its innermost parts, is described to first
order by the Shakura-Sunyaev $\alpha$-model, both $\nu_1$ and $\nu_2$ follow 
a power-law. If viscosities satisfy relation (\ref{eqn:nu1-nu2 relation}) 
and $\alpha$ is assumed to be constant, their exponent are equal:
\begin{equation} \label{eqn:Sa-Su viscosity 1}
\nu_1=A_{\nu_1}R^{\beta} \qquad {\rm and}  \qquad \nu_2=A_{\nu_2}R^{\beta}.
\end{equation}
Following standard Shakura-Sunyaev disc solutions for 
external regions of an accretion disc \citep{FrankKingRaine2002}, 
we have $\beta=3/4$ and 
\begin{equation} \label{eqn:Sa-Su viscosity 2}
\begin{aligned}
& A_{\nu_1}= 9.14 \times 10^{6}\alpha_{0.1}^{4/5}M_{6}^{1/20}\left( \frac{f_\ed}{\eta_{0.1}} \right)^{3/10}{\rm cm^{5/4}s^{-1}}\\
& A_{\nu_2}= \left(\frac{\nu_2}{\nu_1} \right) A_{\nu_1} = 50~f_{\nu_2}~\alpha_{0.1}^{-2}~A_{\nu_1}.
\end{aligned}
\end{equation}
In equation (\ref{eqn:Sa-Su viscosity 2}),
 $\alpha_{0.1}$ and $\eta_{0.1}$ are the $\alpha$ coefficient and the BH 
radiative efficiency in unit of $0.1$. $f_{\nu_2}$ is tabulated in Table~1 (Lodato \& Pringle 2007).

\section{Warped accretion disc}

\subsection{The angular momentum content of discs: extended versus truncated discs}
\label{subsection:disc content}

The dynamics of a fluid element in a misaligned disc around a spinning BH
is given by the combination of three different motions: the Keplerian rotation 
around the BH; the radial drift, due to radial shear viscosity, and finally the
Lense-Thirring precession, due to the gravitomagnetic field $\mathbf{H}_{\rm gm}$ 
generated by $\mathbf{J}_\bh$ \citep[see, e.g.,][]{Weinberg1972,
ThornePriceMacDonald1986}.
In response to Lense-Thirring induced precession, viscous stresses in the disc 
acts rapidly to produce in the vicinity of the BH
an axisymmetric configuration whereby adjacent fluid elements rotates in the
equatorial plane of the spinning BH. The disc thus warps and the warp
disturbance propagates diffusely \citep{PapaloizouPringle1983} in the disc.

As the Bardeen-Petterson effect modifies the inclination 
of the orbital plane of consecutive infinitesimal rings, then the warped profile of the accretion disc can be described by the specific angular momentum density, $\mathbf{L}$, expressed as
\begin{equation} \label{eqn:L definition}
  \mathbf{L}=L\hat{\bf{l}}=\Sigma\Omega_{\rm K} R^2 \hat{\bf{l}}
\end{equation}  
where $\hat{\bf{l}}(R)$ is a unit vector indicating the local direction of the orbital angular momentum, $L$ is the modulus, $\Sigma$ is the surface density of the disc and $\Omega_{\rm K}$ the local Keplerian angular velocity.
The angle describing the tilted disc is defined as 
\begin{equation}
\theta(R)=\cos^{-1} (\hat{\bf{ l}}(R)\cdot \hat{\bfj}_\bh),
\end{equation}
so that ${\hat {\bf l}}(R)$ carries information of the warped structure of the
accretion disc.
The angular momentum of the accretion disc within radius $R$ is given by
\begin{equation} 
\label{eqn:j vector disc within R}
\mathbf{J}_{\rm disc}(R)=\int_{R_{\rm ISO}}^{R} 2 \pi x \mathbf{L}(x)dx
\end{equation}
where the integration domain extends from the innermost stable 
orbit $R_{\rm ISO}$ out to $R$.
In order to calculate the {\it total} disc angular momentum we define 
an outermost radius, $R_{\rm out}$. 
For an extended disc with $R_\out\to \infty$, the 
disc angular momentum $\bfj_{\rm disc}$ always dominates over $\bfj_\bh.$

Real discs are likely to be truncated by their own self-gravity 
that becomes important at distances   
where the disc mass $M_{\rm disc}(R)\sim (H/R)M_\bh$
\citep[see, e.g.,][]{Pringle1981, FrankKingRaine2002, Lodato2007}. 
Outside the truncation radius, gas can be either turned into stars or expelled by winds from stars 
which do form \citep{Levin2007, KingPringle2007}. 
Thus, we are led to define a disc outer edge as the distance where 
the Toomre parameter for stability, 
$Q=\kappa c_s/(\pi G \Sigma)$  
(where $\kappa^2= R(d \Omega^2/dR)+ 4 \Omega^2$), becomes less than unity, and 
the cooling timescale of the clumping gas is less than its dynamical timescale.
When the Toomre parameter drops toward unity, the disc becomes
unstable on a lenghtscale $\lambda = c_s^2/(G\Sigma)$ 
\citep{Polyachenkoetal1997,Levineetal2008};
for a nearly Keplerian, Shakura-Sunyaev $\alpha$-disc, this scale is much
smaller than the disc radial dimension,  and the cooling time of the associated
perturbation is less or of the same order of its orbital period.
Then, as long as the accretion disc can be described as a Shakura-Sunyaev disc\footnote{This condition is fullfilled only for 
$(M_6 f_\ed)/(\alpha_{0.1}\, \eta_{0.1}) \gtrsim 4.3$. 
If this condition is not satisfied
the gas temperature drops below $\sim 10^4$ K, in 
the external region of the disc where $Q$ is still greater than unity.
The change in the opacity likely modifies the structure of the outer disc, 
and we can not explicitly use (\ref{eqn:r out}). 
In this paper we assume that the outer region is sufficiently extended to
provide matter and angular momentum to the inner regions and use
self-consistently  the Shakura-Sunyaev model to describe the disc in regions  where the 
gravitomagnetic interaction takes place. 
},
the external radius can be defined from the condition $Q(R_{\rm out})=1$, so that
\begin{equation}
\label{eqn:r out}
R_{\rm out}=1.21 \times 10^5 \alpha_{0.1}^{28/45} M_6^{-52/45} \left( \frac{f_{\rm Edd}}{\eta_{0.1}} \right)^{-22/45} R_{\rm S},
\end{equation}
where $R_{\rm S}=2GM_\bh/c^2$ is the Schwarschild radius.
At the outer edge of the disc, $\hat{\bf l}(R_\out)=\hat\bfj_{\rm disc, out}$ and $\theta(R_\out)=\theta_{\out}.$

Definitions (\ref{eqn:L definition}) and (\ref{eqn:j vector disc within R}) 
for $\mathbf{L}$ and ${\bfj_{\rm disc}}(R)$ hold for any disc profile. 
At first order, we can neglect details about the warped disc structure 
around $R_\w$ assuming
$\hat{\mathbf{l}} \approx (0,0,1)$, and estimate
the modulus of the orbital angular momentum within radius $R$,
$J_{\rm disc}(R)$, using Shakura-Sunyaev solutions for a flat
disc. In this approximation, the surface density is 
$\Sigma_{\rm flat} \approx \dot{M}/(3 \pi \nu_1)$ 
\citep[see, e.g.,][]{Pringle1981, FrankKingRaine2002} and  
\begin{equation} \label{eqn:L flat}
L(R) \approx \frac{\dot{M}}{3 \pi \nu_1} \sqrt{G \mbh R}.
\end{equation} 
Using equations 
(\ref{eqn:j vector disc within R}) and (\ref{eqn:L flat}), 
and espression ($\ref{eqn:Sa-Su viscosity 1}$) for $\nu_1$ in the case of 
$\beta=3/4$, the modulus of the disc angular momentum within $R$ reads:
\begin{equation}
\label{eqn:j disc R}
J_{\rm disc}(R)=\frac{8}{21}\frac{\dot{M}\sqrt{GM_{\rm BH}}}{A_{\nu_1}}R^{7/4}.
\end{equation}
If espression (\ref{eqn:j disc R}) is estimated at the outer radius 
(\ref{eqn:r out}), the resulting dimensionless ratio
between the disc  and BH angular momenta is
\begin{equation}
\label{eqn:ratio between momenta}
\frac{J_{\rm disc}(R_{\rm out})}{J_{\rm BH}} = 7.3 \, \alpha_{0.1}^{13/45}M_6^{-37/45}\left( \frac{f_{\rm Edd}}{\eta_{0.1}}  \right)^{-7/45}a^{-1}.
\end{equation}
 
\subsection{Timescales and warp radius}
\label{subsection:disc basic equations}

The time-dependent evolution of the disc is described by the continuity equation
\begin{equation} \label{eqn:continuity}
R\frac{\partial \Sigma}{\partial t}+\frac{\partial}{\partial R}\left(v_{\rm R} \Sigma R \right)=0,
\end{equation}
where $v_{\rm R}$ is the radial component of the velocity vector,
and by the equation of conservation of angular momentum.
In presence of a gravitomagnetic field, for a geometrically thin disc characterized by the two viscosities $\nu_1$ and $\nu_2$, the equation reads \citep{Pringle1992}:
\begin{equation} \label{eqn:angular momentum}
\begin{aligned}
& \frac{\partial\mathbf{L}}{\partial t}= - \frac{1}{R}\frac{\partial}{\partial R}(R \mathbf{L} v_{\rm R})+\frac{1}{R}\frac{\partial}{\partial R}\left(\nu_1 \Sigma R^3 \frac{d\Omega}{dR}~ \mathbf{\hat l} \right)+ \\
& \qquad +\frac{1}{R}\frac{\partial}{\partial R}\left(\frac{1}{2}\nu_2 R L \frac{\partial \mathbf{\hat l}}{\partial R} \right) 
+ \frac{2G}{c^2} \frac{\mathbf{J}_{\bh} \times \mathbf{L}} {R^3}. 
\end{aligned}
\end{equation}
The last term is the Lense-Thirring precession term and the associated angular
velocity is
\begin{equation} \label{eqn:omegalt}
\boldsymbol{\Omega}_{\rm LT}(R)=\frac{2G}{c^2}\frac{\mathbf{J}_{\bh}}{R^3}. 
\end{equation}
The time-dependent equation (\ref{eqn:angular momentum}) describes the radial 
drift of matter and the diffusion of  warping disturbances across the
high-viscosity disc.\\ 
This equation introduces several key scales:
\begin{enumerate} 
\item The viscous/accretion timescale for radial drift, related to the 
angular momentum transport parallel to $\bfj_{\rm disc, out}$, $t_{\rm acc}(R)$. 
It can be seen as the time it takes for a fluid element at $R$
to accrete onto the BH \citep[see, e.g.,][]{Pringle1981}
Considering equation (\ref{eqn:angular momentum}), the balance between 
the advection term and the viscous term proportional to $\nu_1$ 
(both on the right side of equation \ref{eqn:angular momentum}) leads to an 
estimate of the accretion time:
\begin{equation} \label{eqn:t_acc}
t_{\rm acc}(R) \sim {R^2}/{\nu_1} .
\end{equation}
According to equation (\ref{eqn:t_acc}), we can introduce the disc 
consumption timescale $t_{\rm disc}$, a concept useful when considering 
transient, truncated discs, as the accretion timescale at the outer radius: 
\begin{equation}
\label{eqn:disc accretion timescale}
\begin{aligned}
& t_{\rm disc} \sim t_{\rm acc}(R_{\rm out}) = \\
& \qquad 1.71 \times 10^{6} \alpha_{0.1}^{-1/45}M_6^{-11/45}\left(\frac{f_\ed}{\eta_{0.1}}  \right)^{-41/45} {\rm yr}.
\end{aligned}
\end{equation}

\item The timescale for warp propagation, related to the radial diffusion of 
gravitomagnetic perturbations  that 
transport the component
of the disc angular momentum lying in the plane of
the disc; 
this scale is inferred from equation (\ref{eqn:angular momentum}) considering
the term proportional to $\nu_2$,
\begin{equation}
\label{eqn:tbp}
t_{\bp}(R) \sim \frac{R^2}{\nu_2}\sim \left( \frac{\nu_1}{\nu_2} \right) t_{\rm acc} (R).
\end{equation}
The physical interpretation of this timescale has been recently investigated  
by solving numerically  equation (\ref{eqn:angular momentum}) for a thin 
disc \citep{LodatoPringle2006}: starting at $t=0$ with a flat disc misaligned 
relative to the fixed BH spin, 
$t \approx t_{\bp}(R)$ indicates the time it takes for the radial diffusion of the warp to reach radius $R$; on longer timescale, the disc approaches a steady
warped state. 

\item
The characteristic extension of the warp $R_\w$, defined as the distance at 
which the Bardeen-Petterson timescale $t_\bp(R)$ equals the 
Lense-Thirring precession timescale $\Omega_{\rm LT}^{-1}$:
\begin{equation} \label{eqn:Rw implicit}
R_\w=\frac{4GJ_\bh}{\nu_2 c^2}.
\end{equation}
For power-law viscosity model, equations (\ref{eqn:jbh definition}), (\ref{eqn:Sa-Su viscosity 2}) and (\ref{eqn:Rw implicit}) 
give
\begin{equation} \label{eqn:warp radius} 
R_\w=476~\alpha_{0.1}^{24/35}f_{\nu_2}^{-4/7}M_{6}^{4/35}\left( \frac{f_{\ed}}{\eta_{0.1}} \right)^{-6/35}a^{4/7}~R_{\rm S}.
\end{equation}
The warp radius represents the dividing between
the outer region for $R \gg R_\w$, where the disc keeps its original 
inclination, given by $\hat \bfj_{\rm disc,out}$, and the inner region 
for $R \ll R_\w$, where the disc aligns (or anti-aligns) its orbital
angular momentum with the BH spin, ${\bf {\hat l}}\parallel \hat\bfj_\bh$. 
The warp radius fixes also the magnitude of the relevant Bardeen-Petterson 
timescale, which reads 
\begin{equation} \label{eqn:tbp at rbp}
\begin{aligned}
& t_\bp(R_\w)= 33.5~\alpha_{0.1}^{72/35}f_{\nu_2}^{-12/7}M_6^{47/35}\\
& \qquad \qquad \times \left( \frac{f_\ed}{\eta_{0.1}} \right)^{-18/35}a^{5/7}{\rm yr}.
\end{aligned}
\end{equation}
If we define the function
\begin{equation} \label{eqn:psi definition}
\psi(R) \doteq \left| \frac{d \mathbf{\hat l}}{dR} \right|
\end{equation}
and $R_{\bp}$  the radius where the disc is maximally deformed
\begin{equation} \label{eqn:Rbp definition}
\Psi \doteq \psi (R_{\bp})=\max \left( \psi \right), 
\end{equation}
we expect that: 
\begin{equation} \label{eqn:nbp definition}
R_{\bp}=n_{\bp}R_\w
\end{equation}
with $n_{\bp}$ of order unity. $R_{\bp}$ has two important properties: first, if it is the radius where the disc is maximally warped, i.e. where the 
diffusive propagation of vertical perturbations is more significant; second, it provides a reliable estimate
of the distance from the BH where the gravitomagnetic interaction is stronger.
From equation (\ref{eqn:angular momentum})
this interaction is proportional to $(\mathbf{L} \times \mathbf{J}_{\bh})/R^3$:  this
term vanishes in the inner part of the disc ($R\ll R_{\bp}$) since the Bardeen-Petterson effect aligns $\mathbf L$ with $\bfj_\bh$, and also in the outer regions  ($R \gg R_{\bp}$), due to the rapid decline with $R$.
Accordingly, the region near $R_{\bp}$ (or equivalently $R_\w$)  is the only one significantly misaligned with $\mathbf{J}_{\bh}$.
\end{enumerate}

\subsection{Analytical solutions} 
\label{Sec:analitic solution}

\begin{figure*}
  \begin{minipage}{0.45\textwidth}
     \centering
    \includegraphics[width=82mm]{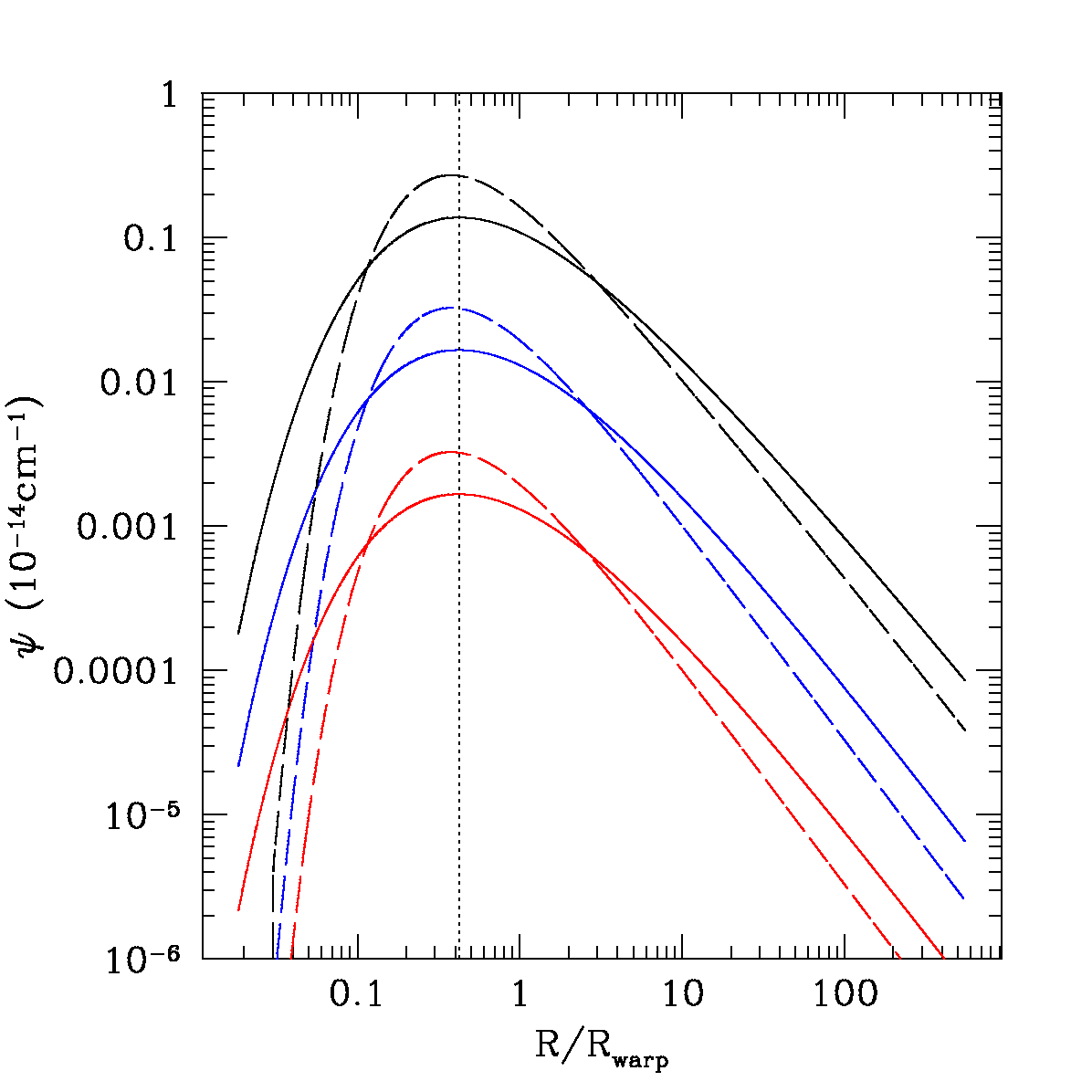}
    \caption{$\psi$ as function of $R/R_\w$, for $W'_{\rm C}$ (solid line) and 
$W'_{\rm PL}$ (dashed line) for different inclination angles: black lines 
correspond to $\theta_\out=\pi /3$, blue lines to $\pi / 30$, red lines 
to $\pi/300$. The vertical black dotted line represents 
$R_{\rm BP}/R_\w=0.42$, the Bardeen-Petterson radius for constant viscosity 
profiles. The parameters set for the BH and the disc is given by 
$\mbho=10^6 \msun$, $a=0.5$, $f_\ed=0.1$, $\alpha=0.09$}
    \label{fig:psi}
    \end{minipage}
\hspace{10mm}
  \begin{minipage}{0.45\textwidth}
    \centering
    \includegraphics[width=82mm]{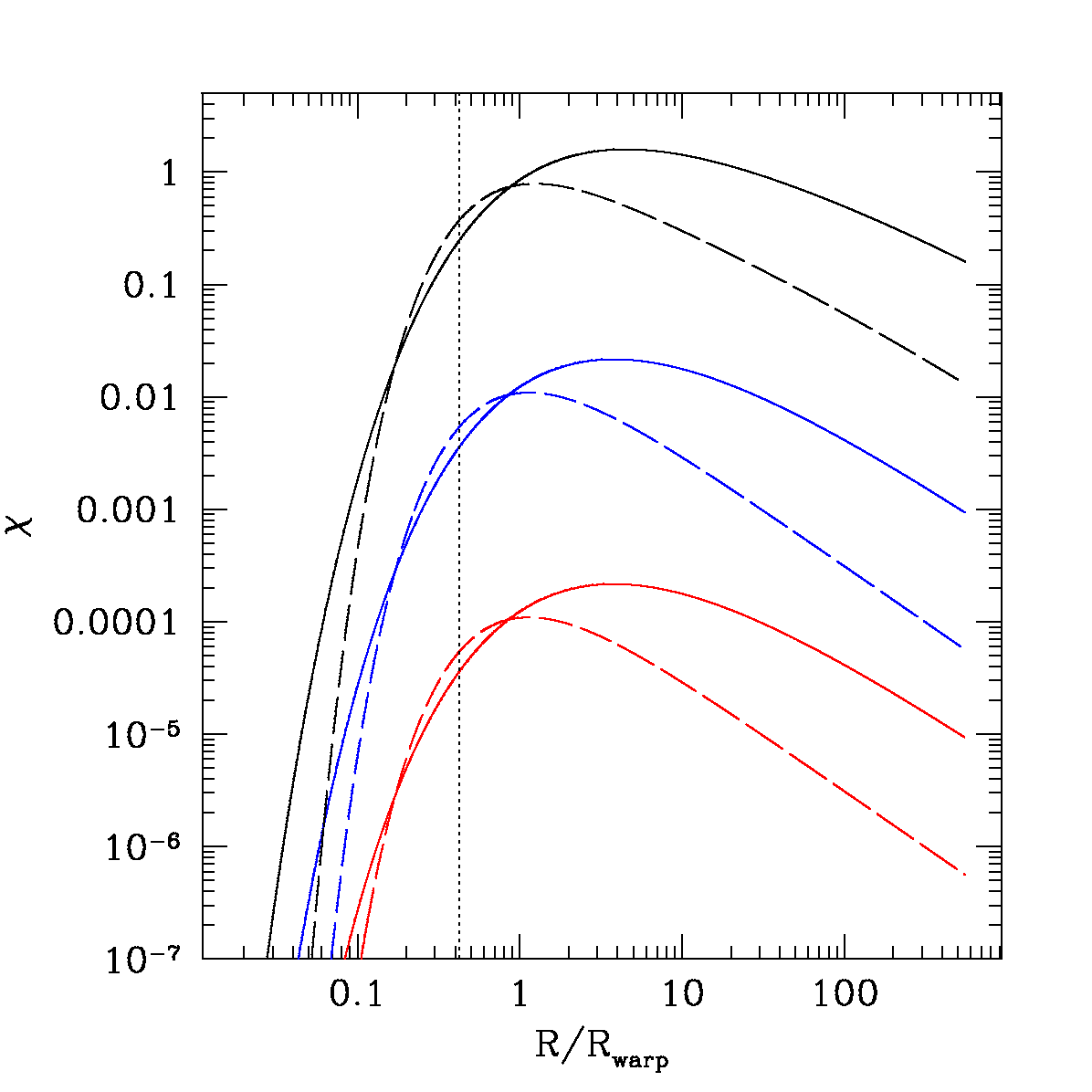}
    \caption{$\chi_{\beta=0}(R)$ (solid line) and $\chi_{\beta=3/4}$ (dashed 
lines) as functions of $R/R_\w$ for 
different inclination angles. 
Black lines correspond to $\theta_\out=\pi /3$, blue ones to $\pi / 30$, 
red ones to $\pi/300$; the vertical black dotted line represents 
$R_{\rm BP}/R_\w=0.42$, the Bardeen-Petterson radius for constant viscosity 
profiles. The parameters set for the BH and the disc is given by 
$\mbho=10^6 \msun$, $a=0.5$, $f_\ed=0.1$, $\alpha=0.09$}
    \label{fig:chi}
  \end{minipage}
\end{figure*}

In this Section we summarise the  properties of the steady warped disc structure 
used to compute the joint evolution of the disc and the BH.

Following previous studies we assume that the viscosity profiles are 
power-laws with exponent $\beta$, as in equation (\ref{eqn:Sa-Su viscosity 1}), and explore two possible cases.
In the first, we formally extend the Shakura-Sunyaev solution 
everywhere in the disc, i.e. $\nu_1\propto R^{\beta}$ with 
$\beta=3/4$, and $\nu_2$ given by equation (\ref{eqn:nu1-nu2 relation}) 
\citep{MartinPringleTout2007}.
In the second case, we assume the viscosities to remain approximately 
constant everywhere in the disc \citep{ScheuerFeiler1996}. 
In order to compare the two models \citep[cfr.]{MartinPringleTout2007}, we impose
the continuity of the viscosities at $R_\bp$  where the gravitomagnetic torque 
is most important.\\
Before solving equations (\ref{eqn:continuity}) and (\ref{eqn:angular momentum})
we introduce two appropriate reference frames. 
The first is the inertial reference frame $Oxyz$ referred to the outer disc; we can always rotate it so that its $z$ axis is parallel to the direction of ${\hat \bfj}_{\rm disc,out}$. 
The second reference frame is the not-inertial frame $O'x'y'z'$ referred to the BH
spin, which is always centered to the BH and whose $z'$ axis is parallel to the black hole time varying spin $\bfj_\bh$ . If we use the adiabatic approximation, then frame $O'x'y'z'$ can be approximated, with time $t$, as a 
sequence of frames, one for every quasi-stationary state of the system. 
The shape of the warped accretion disc is studied in the $O'x'y'z'$ frames and 
the cartesian components of any vector $\boldsymbol{v}$ are there indicated 
as $v'_x,v'_y,v'_z$; $Oxyz$ is the natural frame to study the temporal 
evolution of the BH spin and here the Cartesian components of the previous 
vector are denoted as  $v_x,v_y,v_z$.\\
For a stationary state, continuity equation (\ref{eqn:continuity}) can be 
easily intergrated introducing the accretion rate $\dot{M}$ as 
constant of integration:
\begin{equation} \label{eqn:integral of continuity}
R \Sigma v_{\rm R} = -\frac{\dot{M}}{2 \pi}
\end{equation}
while the projection of equation (\ref{eqn:angular momentum}) along 
$\hat{\mathbf{l}}$ reads:
\begin{equation} \label{eqn:angular momentum 3}
\left(\frac{3}{2}\nu_1\frac{dL}{dR}-\frac{\dot{M}\sqrt{G\mbh}}{4 \pi \sqrt{R}} \right)+\frac{1}{2}R\nu_2 L \left|\frac{d \mathbf{\hat l}}{dR} \right|^2=0
\end{equation}
In the small deformation approximation \citep{ScheuerFeiler1996} 
the warp is gradual and we can neglect the non-linear term, proportional to $| \partial \mathbf{l} / \partial R |^2$. Using the boundary condition $\Sigma(R_{\rm ISO})=0$, the integral of (\ref{eqn:angular momentum 3}) is
\be \label{eqn:solution for L}
L(R)=\frac{\dot M}{3 \pi \nu_1}\sqrt{G M_\bh R} \left( 1-\sqrt{\frac{R_\ISO}{R}}  \right).
\ee
This means that, in this approximation scheme, 
the modulus of the angular
momentum density for a warped accretion disc far from the horizon is 
the same as for a flat disc, equation (\ref{eqn:L flat}).
 
Following \citet{ScheuerFeiler1996} we study the disc profile of the steady 
disc introducing the complex variable $W'={\hat l}'_x+i{\hat l}'_y$ and
considering the case $\theta_{\rm out}< \pi/2$. Using power-law viscosities 
according to (\ref{eqn:Sa-Su viscosity 1}), 
analytic solutions of equation (\ref{eqn:angular momentum}) 
in the small deformation approximation 
have been found by \citet{MartinPringleTout2007}:
\begin{equation} \label{eqn:solution for W, nu power-law}
\begin{aligned}
& W'_{\rm PL}=B \left( \frac{R}{R_\w} \right)^{-\frac{1}{4}}\\
& \qquad \times K_{{1}/{2(1+\beta)}}\left(\frac{\sqrt{2}(1-i)}{(1+\beta)}\left( \frac{R}{R_\w} \right)^{-\frac{1+\beta}{2}} \right)
\end{aligned}
\end{equation}
where $B$ is a complex constant of integration, depending on the boundary condition at the external edge, the subscript "PL" is a reminder of the power-law viscosities and $K_{1/(2(1+\beta))}$ is the modified Bessel function of order $1/(2(1+\beta))$.
In the particular case where we consider constant viscosities, i.e. $\beta=0$, 
the solution can be written as
\begin{equation} \label{eqn:solution for W, nu constant}
W'_{\rm C}=A \exp{\left(-\sqrt{2}~(1-i)\left( \frac{R}{R_\w}\right)^{-\frac{1}{2}} \right) }
\end{equation}
where $A$ is a complex constant of integration and the subscript "C"  is a reminder of the constant viscosity model \citep{ScheuerFeiler1996}. In this latter 
case, $n_{\rm BP}$ is calculated self-consistently, using the definition 
(\ref{eqn:nbp definition}) and the prescription for constant viscosity evaluation at $R_{\rm BP}$;
we find that, for every possible parameters set, $n_\bp \approx 0.42$.
We notice that $W'$ (and so also $\psi$) depends on the radius $R$ through the
dimentionless ratio $R/R_\w$ \citep{MartinPringleTout2007}.

In Figure \ref{fig:psi} we plot the modulus of the gradient of 
$\hat{\mathbf{l}}$, $\psi(R)$, which is a local measure of the deformation 
degree of the disc, for a particular set of parameters and for three different 
angles, $\theta_\out= \pi /3,\pi /30,\pi / 300 $. 
The shape of $\psi$ is similar for the two different disc profiles and for all 
the angles: there is a well defined maximum near $R_\w$, where we expect 
the disc to be more deformed.
At radii smaller than $R_\w$ and far from $R_\w$ the disc is almost flat
(note that the graph is logarithmic in both axes).
For the constant viscosity (power-law) profile the peak is at 
$R_\bp \approx 0.42 R_\w$  ($R_\bp\approx 0.38 R_\w$).
In Figure 1 we also see that a constant viscosity disc is less warped 
(since the maximum deformation is the smaller) than the power law viscosity disc. 
The ratio between the maximum deformations in the power law vs constant viscosity is roughly a factor 2,
and it does not depend on the inclination angle (except for a scale factor, approximately equal to the ratio between the corresponding angles).

\subsection{Validity of the approximation}

\begin{figure*}
  \centering
     \includegraphics[width=144mm]{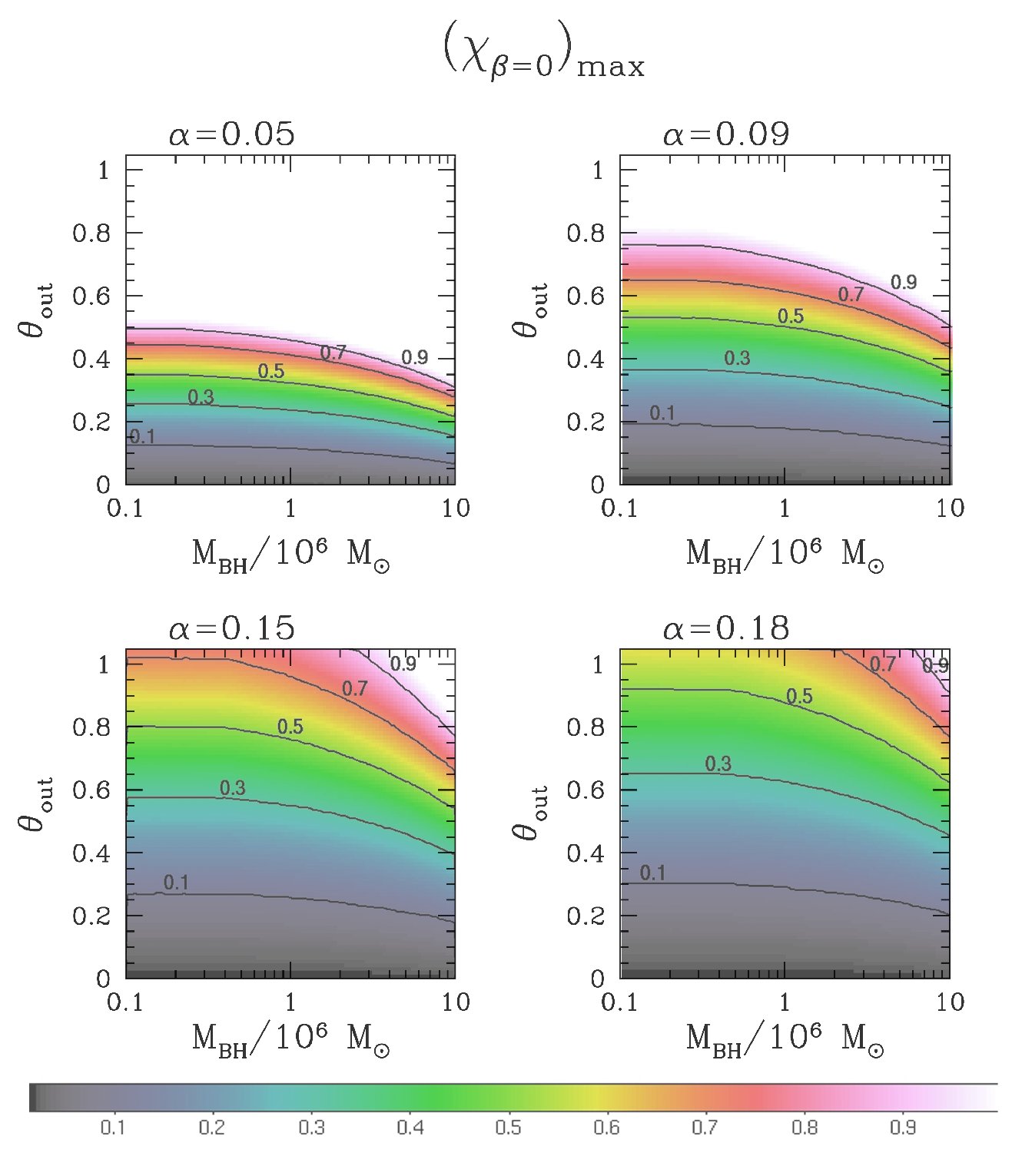}
    \caption{Color coded plot of $(\chi_\beta)_{\rm max}$ in the $\theta_\out$ versus $M_\bh$ plane, for four different $\alpha$ parameters: $\alpha=0.05$ top left panel, $\alpha=0.09$ top right panel, $\alpha=0.15$ bottom left panel, $\alpha=0.18$ bottom right panel. The disc has constant viscosity profiles, i.e. $\beta=0$. The accretion rate is $f_\ed=0.1$ and the spin parameter is $a=0.9$}
    \label{fig:chimaxalpha}
\end{figure*}

\begin{figure*}
  \centering
     \includegraphics[width= 144mm]{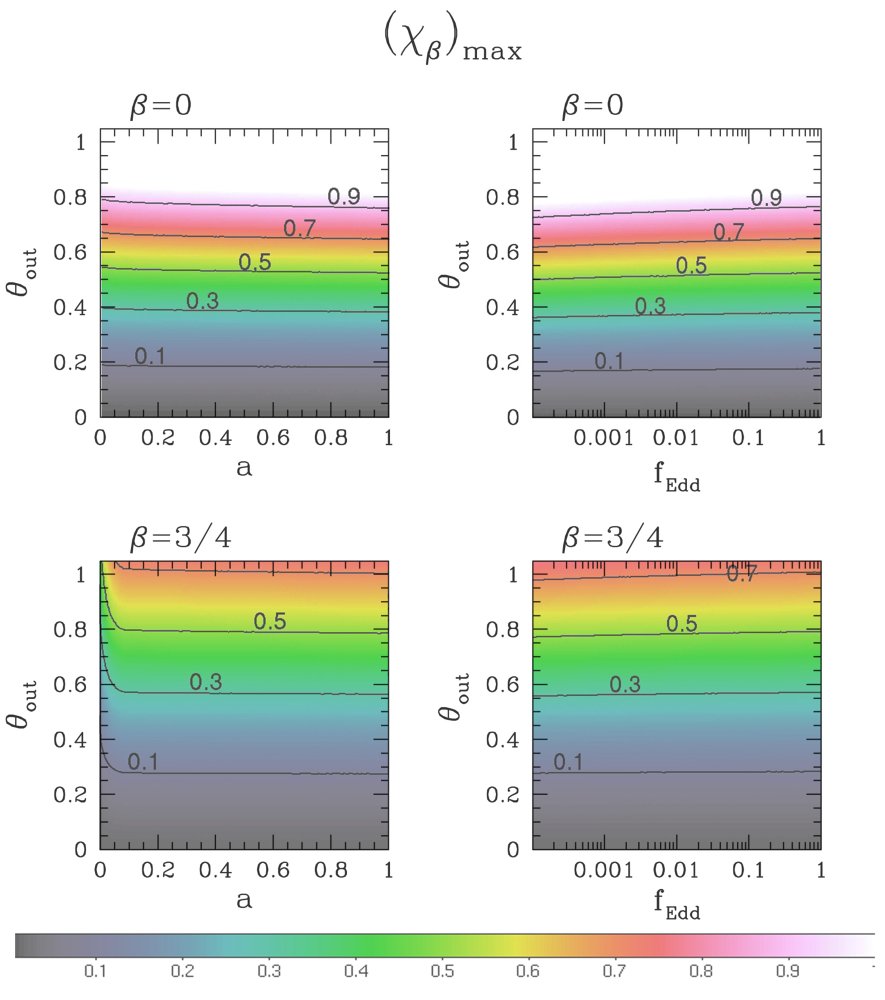}
    \caption{Left panels: color coded plot of $(\chi_\beta)_{\rm max}$ in the 
$a$ versus $\theta_\out$ plane, for $\alpha=0.09$, $\mbh= 10^5 \msun$ and
$f_\ed=0.1$. Right panels: color coded plot of 
$(\chi_\beta)_{\rm max}$ in the $f_\ed$ versus $\theta_\out$ plane, for $\alpha=0.09$, 
$\mbh= 10^5 \msun$ and
$a=0.9$. Top panels refer to constant viscosity profiles, bottom panels to 
power-law viscoisty profiles.}
    \label{fig:chimaxtsd}
\end{figure*}

We calculated the warped disc profile under the small 
deformation approximation. We neglected second order terms in equation 
(\ref{eqn:angular momentum 3}) and found an analytic solution 
for $\mathbf{L}$; in order to verify the consistence of this approximation, 
we define $\chi_{\beta}$ as the ratio between the neglected term and the 
first term into the round brackets of (\ref{eqn:angular momentum 3}), assuming
to have a Keplerian disc with power-law viscosity profile with exponent $\beta$, like in equation (\ref{eqn:Sa-Su viscosity 1}).
Considering equations (\ref{eqn:solution for L}) for $L$, 
we have $dL/dR \approx (1/2+\beta)(L/R) $ and then $\chi_{\beta}$ reads
\begin{equation} \label{eqn:chi definition}
\chi_{\beta}(R)=\frac{2}{3\left(\beta+1 \right)}\frac{\nu_2}{\nu_1}\left|\frac{d\mathbf{l}}{dR} \right|^2 R^2.
\end{equation}
Once we know the explicit solutions, the consistence of this approximation 
can be tested a posteriori calculating $\chi_\beta$: the approximation is well 
satisfied if $\chi_\beta\ll 1.$  
From equation (\ref{eqn:chi definition}), $\chi_\beta$ can be expressed also as
a function of $R/R_\w$ and $\psi$:
\begin{equation}
\chi_\beta =\frac{2}{3\left(\beta + 1\right)}\,\frac{\nu_2}{\nu_1}\,R_\w^{2}\left(\frac{R}{R_\w}\right)^2\psi^2 \left( \frac{R}{R_\w}\right). 
\end{equation} 
Figure \ref{fig:chi} shows the function $\chi_{\beta=0}$  for constant viscosity
profiles (dashed lines) and the function $\chi_{\beta=3/4}$ 
for power-law viscosity profiles (solid lines), for the 
same parameters as in Figure \ref{fig:psi}.\\
The function $\chi_{\beta}$ exhibits a maximum, $(\chi_{\beta})_{\rm max}$, 
around $R_\w$. Far from $R_\w$ the accuracy of the approximation increases, albeit slowly. 
The function $\chi_{\beta}$ is most sensitive to the inclination angle, as expected (notice that Figure 2 uses logarithmic axes).

In Figure \ref{fig:chimaxalpha},  we test the validity of the
small deformation approximation plotting, 
in the BH mass versus $\theta_\out$ plane, the color coded
values of  $(\chi_{\beta=0})_{\rm max}$, for different values of the viscosity parameter 
$\alpha$  (Table \ref{tab:viscosita' lodato}), using the constant 
viscosity profile model (we fix $f_\ed=0.1$ and $a=0.9$).  
White zones represent the regions where $(\chi_{\beta=0})_{\rm max} > 1$, i.e.
where the small deformation approximation becomes invalid.
$(\chi_{\beta=0})_{\rm max}$ shows mainly a strong dependence on inclination angle 
$\theta_\out$, but also a weaker dependence on the BH mass which reveals 
that the small deformation approximation is less accurate for 
$\mbh \gtrsim 10^6 \msun$ and increasing BH mass. 
Comparing different $\alpha$ values, the approximation is better satisfied 
for large viscosities parameters (i.e. $\alpha=0.18$).  
We repeated the analysis for the power-law viscosity model
that shows no significant differences in the parameters dependence.\\
In Figure \ref{fig:chimaxtsd}, using the same colours conventions, we explored  $(\chi_{\beta})_{\rm max}$
in  the $\theta_\out$  versus $a$  (left panels), and $f_\ed$ versus  $\theta_\out$ (right panels) planes, once we have fixed the viscosity parameter ($\alpha=0.09$), the BH mass ($\mbh= 10^6 \msun$), and  $f_\ed=0.1$ for the left panels and $a=0.9$ for the right panels. For both constant ($\beta=0$) and power-law ($\beta=3/4$) models the relative inclination angle is again the leading parameter gauging
the goodness of the fit as the approximation depends very weakly on $a$
and $f_\ed$.

\section{Black hole evolution}

\subsection{Basic equations}
\label{subsection:bh basic equations}

In this section we explore the equations for the BH evolution.
The BH is accreting and its mass increases, from an initial value $M_{\bh,0},$
according to
\begin{equation} \label{eqn:mass evolution}
\frac{d\mbh}{dt}=\dot{M}\frac{E(R_\ISO)}{c^2}
\end{equation}
where $E(R_\ISO)$ is the energy per unit mass of a test 
particle at the innermost
stable orbit.
${E(R_\ISO)/c^2}=1-\eta(a) $ is related to the efficiency $\eta(a)$
that depends only on the spin parameter \citep{Bardeen1970,Bardeenetal1972}.
Equation (\ref{eqn:mass evolution}) introduces a natural timescale for BH
mass growth, known 
as Salpeter time $t_{\rm S}$:
\begin{equation} \label{eqn:Salpeter time}
t_{\rm S}= 4.5 \times 10^{8}\frac{\eta}{f_\ed(1-\eta)}\quad {\rm yr}.
\end{equation}

As argued by \citet{Rees1978} and shown by \citet{ThornePriceMacDonald1986},
there is a coupling between the BH spin and the angular momentum of the disc.
Even though the disc is much less massive that the BH, the moving 
fluid elements perturb the Kerr metric and interact with the BH spin, causing
spin {\it precession,} and if viscous dissipation is present, {\it alignment}.
For an infinitesimal ring of inviscid matter with total angular momentum $\mathbf{J}_{\rm ring}$, the BH spin precesses, following the equation
\begin{equation} \label{eqn:jbh precession-ring}
\frac{d\mathbf{J}_\bh}{dt}=\frac{2G}{c^2}\frac{\mathbf{J}_{\rm ring}}{R^3} \times \mathbf{J}_\bh, 
\end{equation}
with a precession frequency 
\begin{equation}
\Omega^{\rm precession}_\bh=\Omega_{\rm LT} \frac{J_{\rm ring}}{J_\bh}.
\end{equation}
Equation (\ref{eqn:jbh precession-ring}) can be extended to the case of an 
accretion disc to yield:  
\begin{equation} \label{eqn:jbh precession-disc}
\begin{aligned}
&\frac{d\mathbf{J}_\bh}{dt}=\dot{M}\Lambda(R_\ISO)\hat {\bf{l}}(R_\ISO) \\
& \qquad \qquad + \frac{4\pi G}{c^2}\int_{\rm disc}\frac{\mathbf{L}(R) \times \mathbf{J}_\bh}{R^2}dR. 
\end{aligned}
\end{equation} 
The first contribution is due to accretion of matter at $R_\ISO$ 
where $\Lambda(R_\ISO)$ indicates the orbital angular momentum per unit mass
carried by matter at ISO; the Bardeen-Petterson effect ensures that the direction of $
\hat {\bf{l}}(R_\ISO)$ is parallel or anti-parallel to $\hat {\bf{J}}_\bh$, so that the accretion modifies only the spin modulus. 
As shown by \citet{Bardeen1970}, a variation of mass $\Delta \mbh= \sqrt{6}\mbho$ is necessary to pass from a Schwarzschild BH ($a=0$) to an extreme Kerr BH ($a=1$), while spin flip of $\pi$, due only to accretion on an initially extreme Kerr BH, needs $\Delta \mbh= 3 \mbho$. So, the spin accretion timescale for the
spin modulus is of the same order of the mass accretion timescale $t_{\rm S}$.
The second term in equation (\ref{eqn:jbh precession-disc}) describes the 
gravitomagnetic interaction between the rotating viscous disc and the BH spin
vector. 
This term 
modifies only the {\it spin direction} of the BH 
in order to conserve the total angular momentum of the system. 
Under the working hypothesis
that the disc is continually fed by matter carrying the
same angular momentum (see Section \ref{sec:conclusions} for a critical 
discussion), the BH aligns its spin $\bfj_\bh$ in the direction
of ${\hat{\bfj}}_{\rm disc,out}$.  Alignment implies that $\theta_\out(t)=\cos^{-1}
(\hat{\bfj}_\bh(t)\cdot \hat{\bfj}_{\rm disc, out})$ goes to $0$ with 
time.
Figure \ref{fig:integ} shows the function $I$ defined as the 
modulus of the integral kernel of equation (\ref{eqn:jbh precession-disc}) 
\begin{equation}
I(R)= \frac{4 \pi G}{c^2} \frac{L(R)J_{\rm BH} \sin[{\theta(R)}]}{R^2}
\end{equation}
as a function of $R/R_\w,$ for different value of 
$\theta_{\rm out}= \pi / 3, \pi /30, \pi / 300$, where  $\theta(R)$ is
computed along the profile of the steady warped disc of equation
(\ref{eqn:solution for W, nu power-law}) and 
(\ref{eqn:solution for W, nu constant}).
The function $I$, similarly to $\psi$ (defined in eq.  
[\ref{eqn:psi definition}]), peaks near $R_{\w}$.
Contrary to $\psi$, power law viscosity profiles have lower peaks, compared
with constant viscosity profiles.
This figure indicates also that the BH-disc gravitomagnetic interaction 
is spread over a relatively small region of the disc around the warp radius; 
the characteristic spreading length, which is slightly larger for constant 
viscosity profiles, is usually of a few warp radii.

\subsection{Alignment time}

\begin{figure}
  \centering
    \includegraphics[width=82mm]{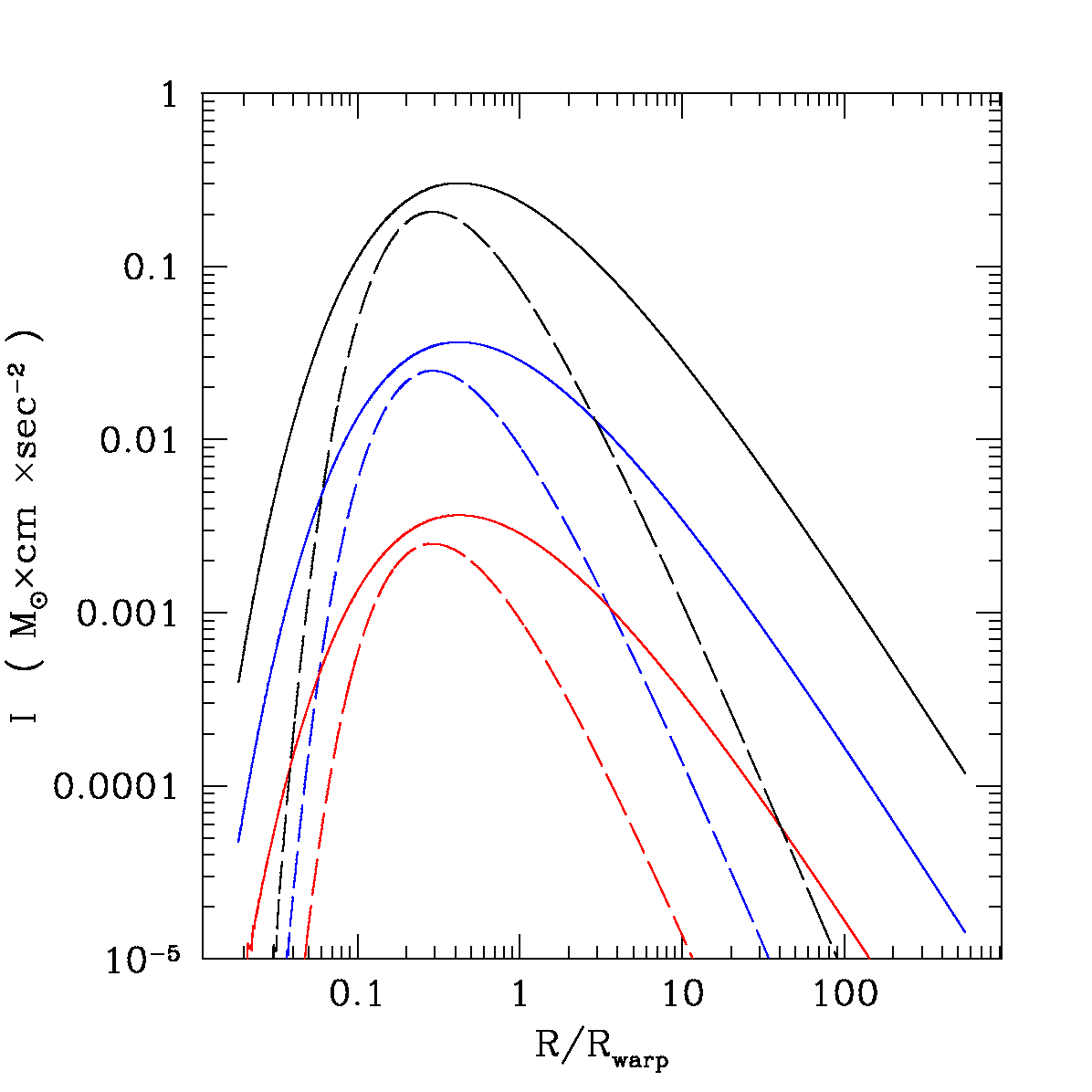}
    \caption{In this figure we draw modulus of gravitomagnetic interaction
term, $I$,  as a function of radius normalized to warp radius, $R/R_\w$. Disc profiles are obtained by BH with $\mbh = 10^6 \msun$ and $a=0.5$, and accretion disc with $f_\ed = 0.1$ and $\alpha=0.09$. Solid (dashed) lines refer to
constant (power-law) viscosity profiles; black lines to $\theta_{\rm out}=\pi / 3$, blue lines to $\theta_{\rm out}= \pi / 30$, red lines to $\theta_{\rm out}= \pi / 300$.}
    \label{fig:integ}
\end{figure}

In this Paragraph we want to give simple estimations for the alignment and
the precession timescales, starting from equation 
(\ref{eqn:jbh precession-disc}).

Assuming BH mass and spin modulus
variations due to accretion to be small compared with
gravitomagnetic effects during the alignment, we neglect the 
term proportional to $\Lambda(R_{\rm ISO})$ in (\ref{eqn:jbh precession-disc});
if BH spin aligns and precess,
left hand side of (\ref{eqn:jbh precession-disc}) can be estimated 
introducing a characteristic gravitomagnetic timescale $\tau_{\rm gm}$ as
\begin{equation}
\nonumber
\left| \frac{d \bfj_{\bh}}{dt}  \right| \sim \frac{\left| \Delta \bfj_\bh  \right|}{\tau_{\rm gm}} \sim \frac{J_{\rm BH}\sin{\theta_\outo} }{\tau_{\rm gm}},
\end{equation} 
and the integral on the right hand side as
\begin{equation}
\nonumber
\left| \frac{4\pi G}{c^2}\int_{\rm disc}\frac{\mathbf{L}(R) \times \mathbf{J}_\bh}{R^2}dR \right| \sim  \frac{4\pi G}{c^2} \frac{L(R_{\rm warp})J_{\rm BH} \sin{\theta_\out}}{R_{\rm warp}}
\end{equation} 
since the bulk of the gravitomagnetic interaction occurs around $R_\w.$
Equating these two expressions and using equation 
(\ref{eqn:solution for L}) for the specific angular momentum density 
modulus, we obtain
\begin{equation}
\label{eqn:T}
\tau_{\rm gm} \sim \frac{3}{4}\frac{c~\nu_1(R_\w)}{G\dot{M}}\sqrt{\frac{R_{\rm warp}}{R_{\rm S}}}.
\end{equation}
Using equation (\ref{eqn:Rw implicit}) and (\ref{eqn:j disc R}) which imply  
$\dot{M}\sqrt{G\mbh}/\nu_1(R_\w) \approx (21/8)\,J_{\rm disc}(R_\w)\,R_\w^{-5/2}$, 
the gravitomagnetic scale $\tau_{\rm gm}$ 
can be written in terms of the Bardeen-Petterson warp timescale (eq. [\ref{eqn:tbp}]):
\begin {equation}
\label{eqn:T-tbp}
\tau_{\rm gm}\sim \frac{4 \sqrt{2}}{7}\, {J_\bh\over J_{\rm disc}(R_\w)}
t_{\rm BP}(R_\w)
\end{equation}
and also in term of the accretion timescale (eq. [\ref{eqn:t_acc}])
\begin {equation} \label{eqn:T-tacc}
\tau_{\rm gm}\sim  \frac{4 \sqrt{2}}{7} \, \frac{\nu_1}{\nu_2} {J_\bh\over J_{\rm disc}(R_\w).}t_{\rm acc}(R_\w)
\end{equation}
where $J_{\rm disc}(R_\w)$ is the disc angular momentum modulus
within the warp radius, estimated by (\ref{eqn:j disc R}).
Finally, considering equation (\ref{eqn:j disc R}) and (\ref{eqn:t_acc}) for $t_{\rm acc}$ 
together with the expression for the spin modulus and Schwarzschild radius,
$\tau_{\rm gm}$ of expression (\ref{eqn:T-tacc}) can be rearranged as
\begin{equation}
\tau_{\rm gm }\sim \frac{3}{2 }\, {a}\, \frac{\nu_1}{\nu_2} \, \frac{\mbh}{\dot{M}}\, \sqrt{{R_{\rm S}\over R_\w}}.
\end{equation}
 
Since the disc carries very little angular momentum at the warp radius, 
from equation (\ref{eqn:T-tbp}) $\tau_{\rm gm} \gg t_\bp,$  always.
The gravitomagnetic BH-disc interaction causes BH spin precession 
and alignment at the same time, and then
introduces two scales related with $\tau_{\rm gm}$, the {\it precession} and 
the {\it alignment} timescales, $t_{\rm prec}$ and  $t_\al$ respectively.
We separate 
their relative importance following \citet{MartinPringleTout2007} results,  
and define the parameter $\mu$, so that
\begin{equation} \label{eqn:2timescales}
t_{\al}=\frac{\tau_{\rm gm}}{\cos{\mu}}, \qquad t_{\rm prec}=\frac{\tau_{\rm gm}}{\sin{\mu}}.
\end{equation} 
The exact value of $\mu$ depends 
on the viscosity profile, and can be estimated either analytically
\citep{MartinPringleTout2007}, or numerically as in this paper. 
Initially, we assume alignment and
precession to have the same timescale, $\cos \mu = \sin \mu = \sqrt{2}/2$
according to \citet{ScheuerFeiler1996}.
Substituting expressions (\ref{eqn:Sa-Su viscosity 2}) for the viscosities, (\ref{eqn:warp radius}) for the warp radius, and (\ref{eqn:T}) for $\tau_{\rm gm}$ in (\ref{eqn:2timescales}), the alignment time 
reads
\begin{equation}
\label{eqn:alignment timescale}
t_\al = 1.13 \times 10^5 \alpha_{0.1}^{58/35}f_{\nu_2}^{-5/7}M_6^{-2/35}
\left(\frac{f_\ed}{\eta_{0.1}} \right)^{-32/35}a^{5/7}~{\rm yr}.
\end{equation}
The timescale $t_\al$ increases with $a$, indicating that a rapidly rotating Kerr BH
offers some resistance before changing its direction.  
Interestingly, the alignment timescale does not depend
on the initial inclination $\theta_\outo$ since a more inclined configuration 
implies more pronunced disc deformations and stronger mutual gravitomagnetic 
interactions (as also shown in 
Figure \ref{fig:psi} and \ref{fig:integ}). 
$t_\al$ has a weak dependence on the BH mass and scales nearly as
${\dot M}^{-1}$: a higher accretion rate implies a higher 
angular momentum density ${\bf L}(R)$ and thus a stronger
gravitomagnetic coupling. 
We notice also that, apart from numerical factors of order unity, 
this timescale is consistent with the alignment scales found by 
by \citet{ScheuerFeiler1996,NatarajanPringle1998, NatarajanArmitage1999,MartinPringleTout2007}.

\subsection{The adiabatic approximation}
\label{subsection:the adiabatic approximation}

\begin{figure*}
  \begin{minipage}{0.45\textwidth}
     \centering
    \includegraphics[width=82mm]{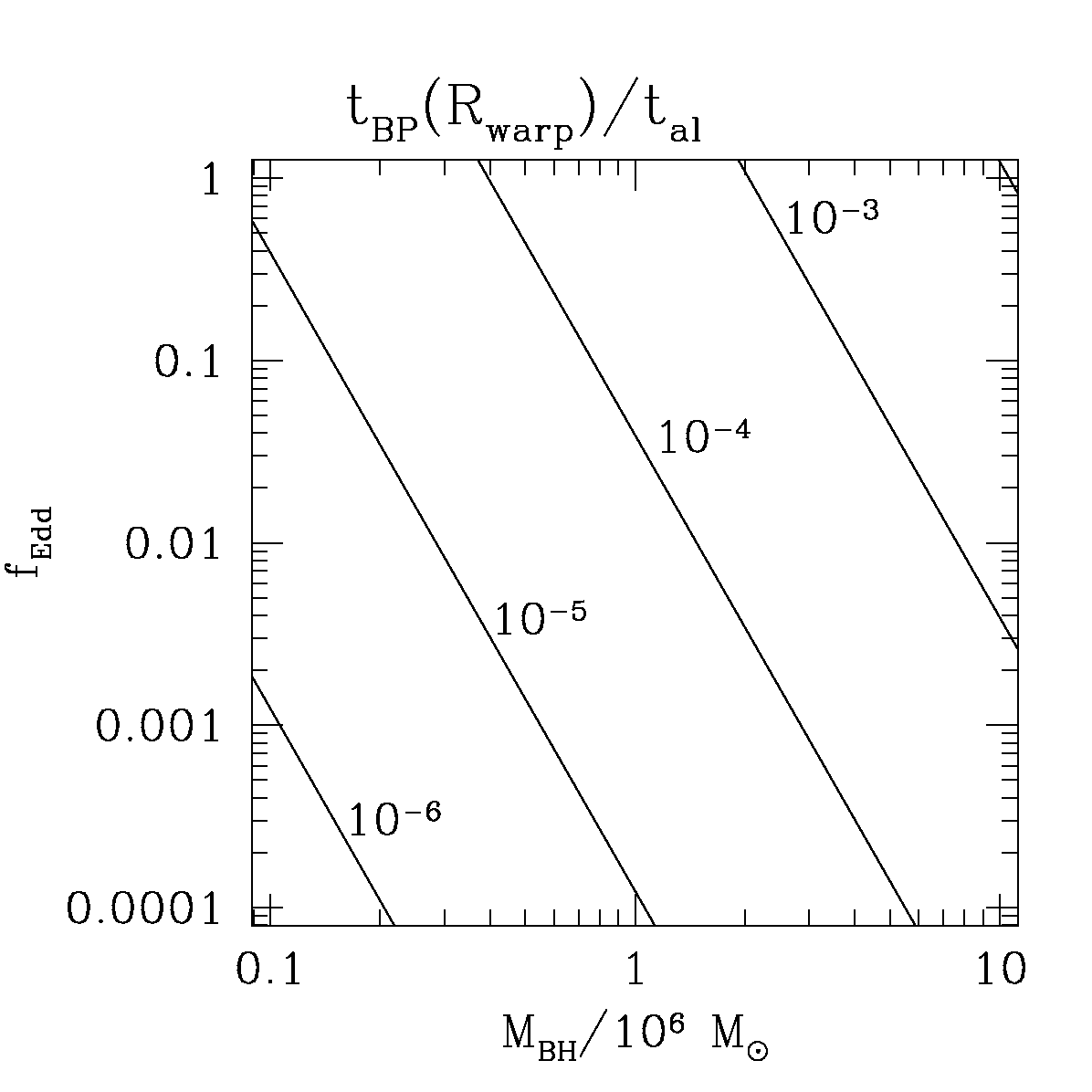}
    \caption{In the $M_\bh$ versus $f_\ed$ plane, we draw lines of constant 
$t_{\rm BP}(R_\w)/t_{\rm al}$ ratio for a BH with $a = 0.9$ and an
accretion disc with $\alpha = 0.09$.}
    \label{fig:twtal}
    \end{minipage}
\hspace{10mm}
  \begin{minipage}{0.45\textwidth}
    \centering
    \includegraphics[width=82mm]{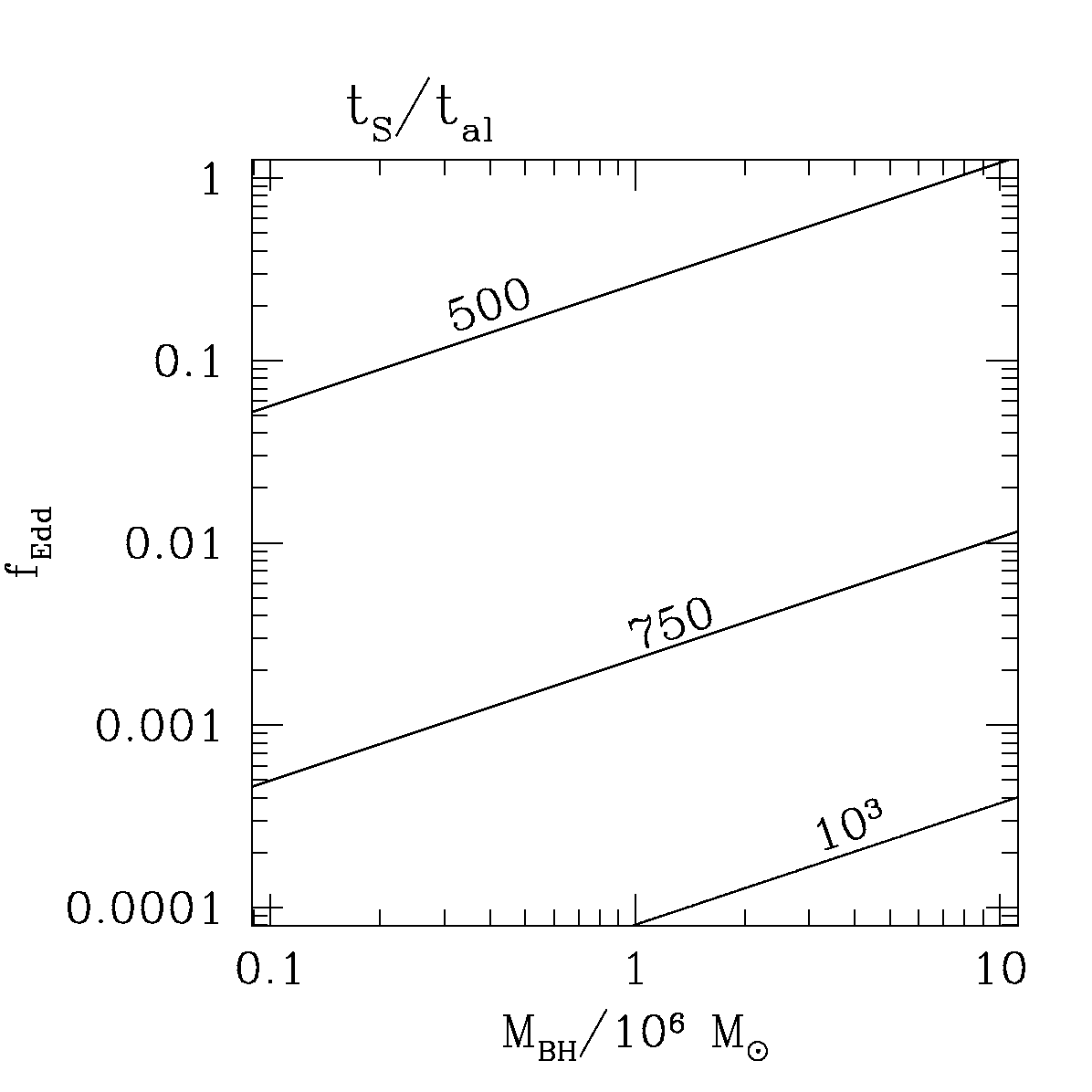} 
    \caption{In the $M_\bh$ versus $f_\ed$ plane, we draw lines of constant 
$t_{\rm S}/t_{\rm al}$ ratio for a BH with $a=0.9$ and an
accretion disc with $\alpha = 0.09$.}
    \label{fig:tstal}
  \end{minipage}
\end{figure*}

In Section \ref{subsection:disc basic equations} and 
\ref{subsection:bh basic equations} we described the equations governing the 
evolution of a warped accretion disc around a fixed BH, and the 
evolution of an 
accreting Kerr BH in gravitomagnetic interaction with its accretion disc.
The BH and the accretion disc evolve contemporary and their evolution is 
coupled, 
so that we can solve simultaneously equations (\ref{eqn:continuity}) and 
(\ref{eqn:angular momentum}) for a Keplerian  disc and 
(\ref{eqn:mass evolution}) and (\ref{eqn:jbh precession-disc}) for the 
accreting and precessing BH.
 
In this paper, we solve these coupled equations using the {\it adiabatic} 
approximation that
separates the rapid temporal evolution of the warped disc from the
longer temporal evolution of the BH.
Equations are integrated starting from given initial conditions: 
at $t=0$ the BH spin $\hat{\bfj}_{\rm BH}$ is inclined with respect 
to $\hat \bfj_{\rm disc,out}$ by an angle $\theta_\outo,$
and the warped disc profile is described by a quasy-stationary profile 
${\bf L}(R,t=0)$; $\mbho$ and $\bfj_{\rm BH,0}$ are the initial 
BH mass and spin.\\
In order to justify this approximation scheme,
we survey  the BH and disc timescales, as functions of $M_\bh$ and $f_\ed$, for
two selected values of the viscosity and spin parameter: $\alpha = 0.15$ and $a=0.9.$  
In Figure (\ref{fig:twtal}) and in Figure (\ref{fig:tstal}), we draw in the 
$M_\bh$-$f_\ed$ plane lines of constant  
$t_\bp(R_w)/t_\al$ and $t_{\rm S}/t_\al$ ratios.
The comparison between the different timescales lead to the following
hierarchy of timescales:
\begin{equation}
t_\bp (R_\w) \ll t_\al\ll t_{\rm S}.
\end{equation}
Then, in the adiabatic approximation, the disc transits through 
a sequence of warped states
over the shortest timescale $t_\bp(R_\w)$ while, on the longer 
timescale $t_\al,$ the BH aligns its spin to $\hat\bfj_{\rm disc, out}$, 
and modifies a little its spin modulus and mass due to accretion. 
Considering one of these disc quasi-steady states, initially at time $t$, 
after a time gap $\delta t \sim t_\bp(R_\w)$ the BH mass and spin $\bfj_\bh$ are updated according to
\begin{equation} \label{eqn:update}
\left\{
\begin{aligned}
&  \mbh(t+t_\bp(R_\w))= \mbh(t)+\delta \mbh  \\
& \mathbf{J}_\bh(t+t_\bp(R_\w))= \mathbf{J}_\bh(t)+\delta \mathbf{J}_\bh 
\end{aligned}
\right.
\end{equation}
and these variations produce a new quasi-stationary warped state at 
$t+t_{\rm BP}(R_\w)$, ${\bf L}(R,t+t_{\rm BP}(R_\w))$.\\
For the BH mass variation $\delta \mbh$, we integrate  equation (\ref{eqn:mass evolution}) 
from $t$ to $t+t_\bp(R_\bp)$:
\begin{equation} \label{eqn:mass variation}
\delta \mbh \approx \dot{M}\frac{E(R_{\ISO})}{c^2}~t_\bp(R_\bp)
\end{equation}
where $R_{\ISO}$ is the last innermost stable orbit associated with the
current  value of $a(t)$.\\
For the spin variation, we need to integrate  equation (\ref{eqn:jbh precession-disc})
that includes 
the two different and coupled contributions due to accretion and gravitomagnetic interaction; if $\delta \mbh$ and $\delta \mathbf{J}_\bh$ are small on the timescale $t_\bp(R_\bp)$, to first order the two contributions decouple and  they can be integrated 
separately: 
\begin{equation} \label{eqn:spin modulus variation}
\left( \delta J_\bh \right)_\acc \approx \dot{M} \Lambda(R_{\ISO})~t_\bp(R_\bp)
\end{equation}
\begin{equation} 
\label{eqn:spin gravitomagnetic variation}
\begin{aligned}
& \left( \delta \mathbf{J}_\bh \right)_{\rm gm} \approx \frac{4 \pi G}{c^2} t_\bp(R_\bp) \\
& \qquad \qquad \times \int_{\rm disc}\frac{\mathbf{L}(R,t) \times \mathbf{J}_\bh(t)}{R^2}~dR 
\end{aligned}
\end{equation}
where $\left( \delta J_\bh \right)_\acc$ is due to accretion and changes 
only the spin modulus while $\left( \delta \mathbf{J}_\bh \right)_{\rm gm}$ 
is due to gravitomagnetic interaction and changes only the spin direction.
After the interval $t_\bp(R_\bp),$ the angular momentum of  (\ref{eqn:update}) 
are updated according to this rule
\begin{equation} \label{eqn:total spin variation}
\begin{aligned}
& \mathbf{J}_\bh(t+t_\bp(R_\w))= \left( \mathbf{J}_\bh(t)+\left( \delta \mathbf{J}_\bh \right)_{\rm gm}\right) \\
& \qquad \times \frac{J_\bh(t)+ \left( \delta J_\bh \right)_\acc}{J_\bh(t)} 
\end{aligned}
\end{equation}
This procedure can be repeated iteratively  on a timescale $t_\al$ to study 
the coupled evolution of $\mathbf{L}(R,t)$, $\mathbf{J}_\bh$ and $\mbh$ during
the alignment process.

\section{Spin alignment}

\begin{figure*}
  \centering
     \includegraphics[width= 144mm]{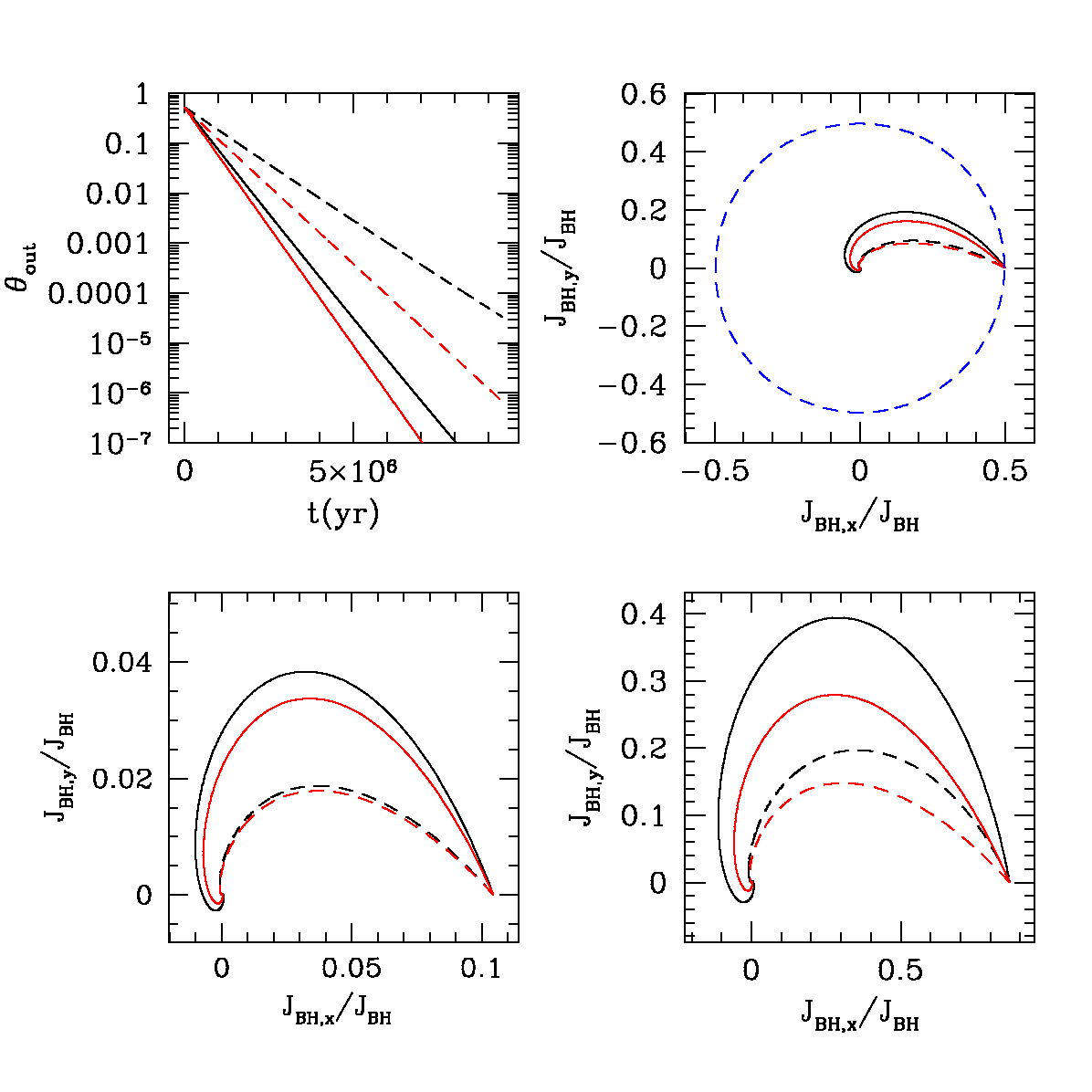}
    \caption{Results for precession and alignment processes. 
Black lines refer to our result while red lines refer to results published by \citet{MartinPringleTout2007}; 
solid lines (dashed lines) refer to constant (power-law) viscosity profile. 
Top left panel represents temporal evolution of relative inclination angle $\theta_\out$ 
while top right shows evolution of $J_{\rm BH,x}/J_\bh$ agains $J_{\rm BH,y}/J_\bh$, both for an initial BH with $\mbho=10^6 \msun$, $a_0=0.5$ and an accretion disc with $f_\ed=0.1$ and $\alpha=0.09$, with $\theta_\outo= \pi / 6$. 
Blue dashed line represents the evolution of the spin components for a 
pure precession motion around $\hat{\bfj}_{\rm disc,out} \, || \, \hat{z}$. 
In bottom left (right) panel we represent evolution of $J_{\rm BH,x}/J_\bh$ agains $J_{\rm BH,y}/J_\bh$ for an initial relative inclination angle $\theta_\outo= \pi / 30$ ($\theta_\outo= \pi /3$), for an initial BH with $\mbho= 10^6 \msun$, $a_0=0.5$ and an accretion disc with $f_\ed=0.1$ and $\alpha=0.09$.}
    \label{fig:spinth}
\end{figure*}

\begin{figure*}
  \centering
     \includegraphics[width=144mm]{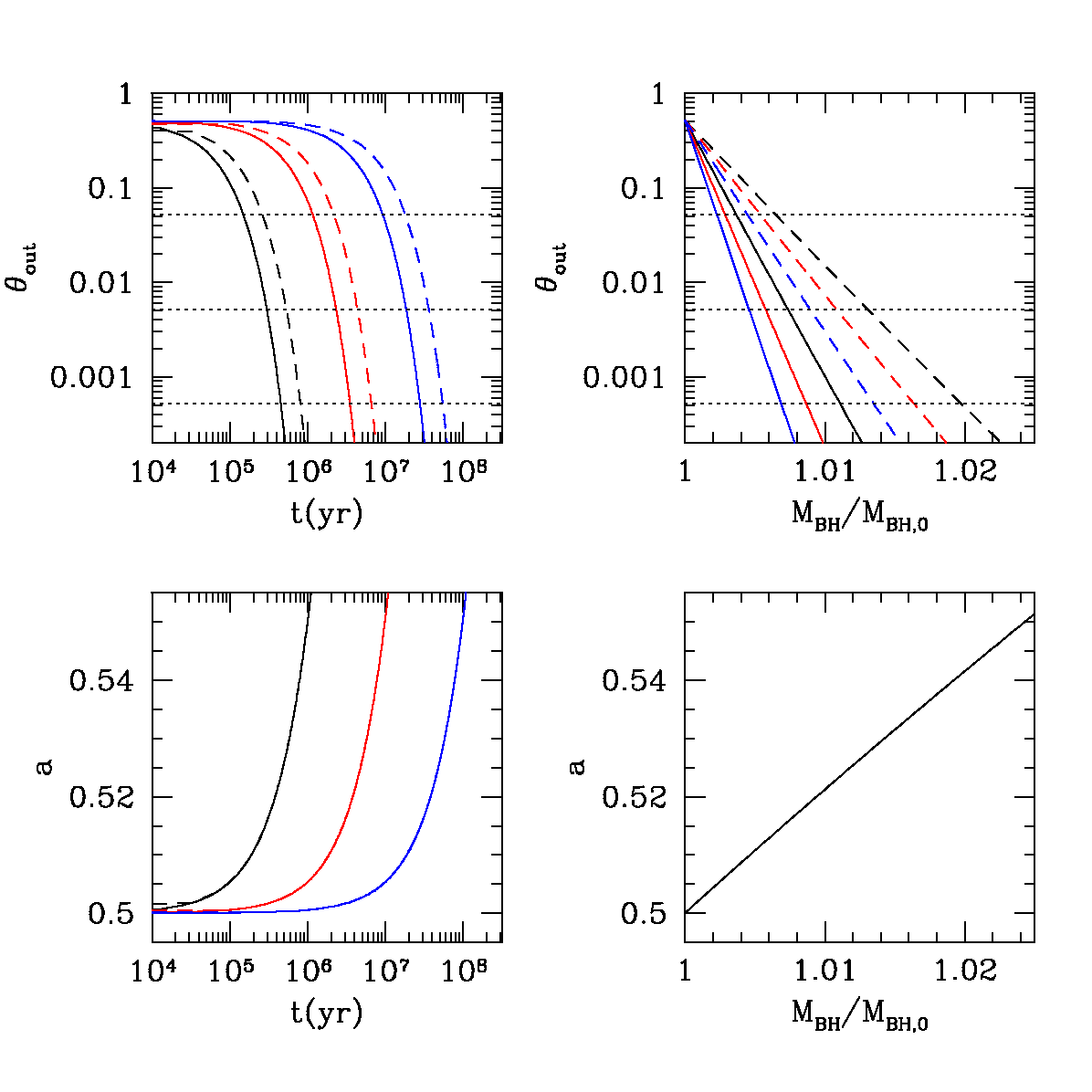}
    \caption{Coupled evolution of the relative inclination angle $\theta_\out$,
BH mass $\mbh$ and spin parameter $a$. Solid (dashed) lines refer to constant 
(power-law) viscosity profile. Black lines to $f_\ed=1$, red lines to 
$f_\ed=0.1$, blue lines to $f_\ed=0.01$. Dotted horizontal lines which appear 
in top panels represent angles $\theta_\out / \theta_\outo=10^{-1},10^{-2},
10^{-3}$. Initial configuration: $\mbho=10^6 \msun$, $a_0=0.5$, $f_\ed=0.1$ 
and $\alpha=0.09$, with $\theta_\outo= \pi / 6$.}
    \label{fig:vsl}
\end{figure*}

\subsection{Set up}

In this Section we study the coupled evolution of the BH and warped accretion 
disc using the approximation scheme described in the previous Section, 
in order to infer
the evolution of $M_\bh$ and $\bfj_\bh$ as a function of time, in response 
to the gravitomagnetic interaction and matter accretion.\\

\begin{table*}
   \begin{center}
     \begin{tabular}{|c|c|c|c|c|c|c|}
       \hline
VP & $f_\ed$  & $\theta_\out / \theta_{out,0}$ & $\Delta t$   & $\Delta t/t_\al $  & $\Delta \mbh/M_{\bh,0}$ & $\Delta a/a_0$ \\
 & &  & ($10^6{\rm yr}$)& &(in units of $10^{-2}$) & (in units of $10^{-2}$)\\
      \hline
      \hline
           &      & $10^{-1}$ & 0.15  & 2.2   & 0.67  & 1.57  \\
           & 1    & $10^{-2}$ & 0.30  & 4.4  & 0.74  & 3.13 \\
           &      & $10^{-3}$ & 0.45  & 6.6  & 1.11  & 4.69 \\
      \hline
           &      & $10^{-1}$ & 1.12   & 2.0   & 0.29 & 1.24 \\
      C    & 0.1  & $10^{-2}$ & 2.33   & 4.2  & 0.58 & 2.46 \\
           &      & $10^{-3}$ & 3.52   & 6.3  & 0.87 & 3.68 \\
     
      \hline
           &      & $10^{-1}$ & 9.28   & 2.0   & 0.23 & 0.99  \\
           & 0.01 & $10^{-2}$ & 18.6   & 4.1  & 0.46 & 1.96  \\
           &      & $10^{-3}$ & 27.9   & 6.1  & 0.68 & 2.94  \\
      \hline
      \hline
           &      & $10^{-1}$ & 0.26   & 3.8   & 0.65 &  2.77 \\
           & 1    & $10^{-2}$ & 0.53   & 7.8   & 1.30 &  5.49 \\
           &      & $10^{-3}$ & 0.81   & 11.9  & 1.98 &  8.20 \\
       \hline
           &      & $10^{-1}$ & 2.18   & 3.9   & 0.54 & 2.30 \\
       PL  & 0.1  & $10^{-2}$ & 4.39   & 7.9   & 1.08 & 4.57 \\
           &      & $10^{-3}$ & 6.64   & 11.9  & 1.64 & 6.84 \\     
      \hline
           &      & $10^{-1}$ & 18.1   & 3.9  & 0.45  & 1.91 \\
           & 0.01 & $10^{-2}$ & 36.3   & 7.9  & 0.89  & 3.79 \\
           &      & $10^{-3}$ & 54.7   & 11.9 & 1.35  & 5.67 \\
      \hline
    \end{tabular}    
     \caption{Summary of our parameters and results for the co-rotating case; we consider viscosity coefficient $\alpha=0.09$ and initial inclination  angle $\theta_\outo=\pi/ 6 $,  both for constant (C) and power-law (PL) viscosity profiles (VP). 
The initial BH has $\mbho = 10^6 \msun$ and $a_{0} = 0.5$.
Accretion rate $f_\ed$ varies over three orders of magnitude and we record times needed to decrease the relative inclination angle of a factor 10, 100 or 1000,
 comparing it with estimated alignment timescale, equation (\ref{eqn:alignment timescale}); we also report mass and spin relative variations.}
   \label{tab:angle variations}
   \end{center}
\end{table*}

At $t=0$ the outer disc, extending up to a radius $R_{\rm out}$, defines  the fixed reference frame $Oxyz$. In this frame the external edge of the disc lies in the $x,y$ plane and the orbital angular momentum at $R_{out}$ is
\begin{equation} \label{eqn:initial disc position}
(L_x(R_\out),L_y(R_\out),L_z(R_\out))=L(R_\out)(0,0,1)
\end{equation}
while the BH spin is initially inclined of $\theta_\outo$ with respect to the $z$ axis: 
\begin{equation} \label{eqn:initial BH spin position}
(J_{\bh,x},J_{\bh,y},J_{\bh,z})=J_\bh (\sin{\theta_\outo},0,\cos{\theta_\outo}).
\end{equation}
If at  $t \neq 0$ we know the components of $\mathbf{J}_\bh$ in the fixed reference frame $Oxyz$ 
there is always a rotated reference frame $Ox'y'z'$ where $\mathbf{J}_\bh$ is along the new $z'$ axis (see also the discussion about reference frames of 
Section \ref{Sec:analitic solution}). 
The two reference frames are related by a rotation $\mathcal{R}$, which 
depends only on the components $J_{\bh,x}$, $J_{\bh,y}$, $J_{\bh,z}$ of 
$\mathbf{J}_\bh(t)$ in the fixed reference frame. 
If $\mathcal{R}_{ij}$ is the matrix associated with this rotation, 
we can easily find the components of $\mathbf{J}_\bh (t)$ and 
$\mathbf{L}(R_\out,t)$ in the rotated frame:
\begin{equation}
\begin{aligned}
& J'_{\bh,i}(t)=\mathcal{R}_{ij}J_{\bh,j}(t)\Rightarrow J'_\bh(t)=(0,0,J_\bh(t))\\
& L'_{i}(R_\out,t)=\mathcal{R}_{ij}L_{j}(R_\out,t). 
\end{aligned}
\end{equation}
As shown by \citet{ScheuerFeiler1996} for the constant viscosity profile and \citet{MartinPringleTout2007} for the power-law viscosity profile, in this special rotated frame of reference it is possible to calculate analytically the expression of the gravitomagnetic torque, using equation (\ref{eqn:spin gravitomagnetic variation}):
 \begin{equation} \label{eqn:torque integral}
\begin{aligned}
& (\delta J'_{\bh,x}+i\delta J'_{\bh,y})_{\rm gm}= \\
& \left( -i \frac{4 \pi G J_\bh(t)}{c^2}\int_{\rm disc}\frac{L(R,t)~W'(R,t)}{R^2}~dR \right)~t_\bp(R_\w)
\end{aligned}
\end{equation}
where $L(R,t)$ is given by (\ref{eqn:solution for L}) and $W'(R,t)$ 
by (\ref{eqn:solution for W, nu constant}) or 
(\ref{eqn:solution for W, nu power-law}). 
The analytic expressions of the gravitomagnetic torques for the two viscosity 
profiles are reported in the Appendix. 
From the torques it is possible to find the values of the spin variations 
$(\delta J'_{\bh,x})_{\rm gm}$, $(\delta J'_{\bh,y})_{\rm gm}$ and 
$(\delta J'_{\bh,z})_{\rm gm}$ in this rotated reference frame. 
Finally, in order to know their expressions in our fixed reference 
frame $Oxyz$, we have to rotate them back, using the inverse rotation 
$\mathcal{R}^{-1}$
\begin{equation}
(\delta J_{\bh,i})_{\rm gm}= (\mathcal{R}^{-1})_{ij} (\delta J'_{\bh,j})_{\rm gm}
\end{equation}
Once we know the spin variations due to gravitomagnetic coupling, the modulus 
variation can be calculated from equation (\ref{eqn:spin modulus variation}) 
and the global spin variation from equation (\ref{eqn:total spin variation}).

\subsection{Results}

We computed, within the adiabatic approximation, the joint evolution of
the BH mass and spin during the process of alignment 
under the assumption that matter is  corotating with the BH. 
We iterated equations (\ref{eqn:mass variation}) and (\ref{eqn:total spin variation}), 
from the initial conditions (\ref{eqn:initial disc position}) and (\ref{eqn:initial BH spin position}), recording the updated values of $M_\bh$,  $a$ and of the
relative inclination angle $\theta_\out$ every snapshot of time 
$\delta t\sim t_\bp(R_\w)$. We initially choose a spinning 
BH with $\mbh=10^6 \msun$ and $a=0.5$, and an accretion disc with 
$\dot{M}=0.1~\dot{M}_\ed$, $\alpha=0.09$; both power-law and constant viscosity profiles are
considered.
Three initial relative inclination angles of $\theta_\outo=\pi/3, \pi/6, \pi/30$ have been tested.
Figure \ref{fig:spinth} shows as a function of time
the inclination angle $\theta_\out(t)$ and
the two components of the BH spin unit vector in the plane $Oxy$;
red lines refers to the analytic solutions given by \citet{MartinPringleTout2007}.
As shown in top-left panel of Figure {\ref{fig:spinth}}, 
the relative inclination angle $\theta_\out$ decreases 
exponentially with time on the scale $t_\al$ and the
decrease is more rapid for the constant viscosity disc (solid line).\\ 
The BH spin aligns with the external disc and precesses, as illustrated in the 
the top-right panel, where we compare also the actual evolution of the spin 
versor with a pure precessional motion (blue dashed line).
Our results are only qualitatively consistent with Martin's results; in our calculations the alignment process appears to be less efficient and
the spin precession more pronounced. The difference between our results and 
Martin's analytical solutions arises from three facts: (i) we included mass and
spin modulus evolution; (ii)  Martin et al. neglected to carry out the rotation connecting  the BH reference frame $O'x'y'z'$ to the disc frame $Oxyz$; 
(iii) for constant viscosity profile, we evaluate $\nu_1$ and $\nu_2$ from
equation (\ref{eqn:Sa-Su viscosity 2}) at the
Bardeen-Petterson radius, $R_{\rm BP}\approx 0.4 R_\w$, while Martin et al. 
evaluate them at the warp radius, $R_\w$. For an initially not very inclined 
BH spin, the difference tends to disappear, because the rotation matrix 
nears the identity matrix.
However, for large $\theta_\outo$ the discrepancy becomes more important  
(see, e.g., bottom panels of Figure \ref{fig:spinth}).

Figure \ref{fig:vsl} shows the evolution of $\theta_\out$ and $a$ 
as functions of time and of the increasing BH mass, for an initial BH with  
$M_{\bh,0}= 10^6 \msun$, $\theta_\outo=\pi/6$ and spin parameter $a_{0}=0.5$,
and for  $f_\ed=1,0.1$, and $0.01$.
Both constant and power-law viscosity profiles are explored, always with viscosity parameter $\alpha=0.09$.
Alignment is a process that shows a strong dependence on the accretion rate:
for the constant (power-law) viscosity model the time necessary to reduce the relative inclination angle by a factor 100 varies 
from $3.0 \times 10^5{\rm yr}$ ($5.3 \times 10^5{\rm yr}$) for $f_\ed=1$ 
to $1.86 \times 10^7{\rm yr}$ ($3.63 \times 10^7{\rm yr}$) for $f_\ed=0.01$. 
During this alignment time,
the BH has increased its mass by a small fraction, 
between $0.74\%$ ($1.30\%$) for $f_\ed=1$ and $0.46\%$ ($0.89\%$) for $f_\ed=0.01$.
The spin parameter $a$ increases due to accretion, but only by a small amount, 
between $3.13\%$ ($5.47\%$) for $f_\ed=1$ to $1.96\%$ ($3.79\%$) for 
$f_\ed=0.01$.\\
In Table \ref{tab:angle variations} we summarize the results of Figure \ref{fig:vsl}. 
There, we compare also the time $\Delta t$ necessary to decrease the initial 
inclination angle by a given amount, with $t_{\rm al}$ estimated from equation (\ref{eqn:alignment timescale}):
the values of the time $\Delta t$  are consistent with the interpretation of  $t_{\rm al}$ as  $e$-folding time. For constant viscosity profiles,
a closer match of $t_{\rm al}$ with the numerical outcomes requires 
$(\cos \mu)_{\rm C} \approx 0.78$ instead of $\sqrt{2}/2$, and 
$(\cos \mu)_{\rm PL} \approx 0.41$ for power-law viscosity profiles
(subscripts C and PL simply remind that for different viscosity prescriptions
we found different $cos{\mu}$ values).
Then, the ratios between the precession and the alignment timescales are
$(t_{\rm prec}/t_{\rm al})_{\rm C}=0.81$ and $(t_{\rm prec}/t_{\rm al})_{\rm PL}=2.2$. 
These results are still qualitatively consistent with \citet{MartinPringleTout2007}, who
have shown that both $t_{\rm al}$ and $t_{\rm prec}/t_{\rm al}$ increase 
with the exponent $\beta$ of the viscosity profile; small quantitative 
differences are due to different assumptions and different calculations 
methods. Finally, we notice that the scaling of $\Delta t$ with $f_\ed$ is in
good agreement with estimation (\ref{eqn:alignment timescale}).

\subsection{Exploring the parameters space}

\begin{figure*}
\begin{minipage}{0.45 \linewidth}
  \centering
     \includegraphics[width=82mm]{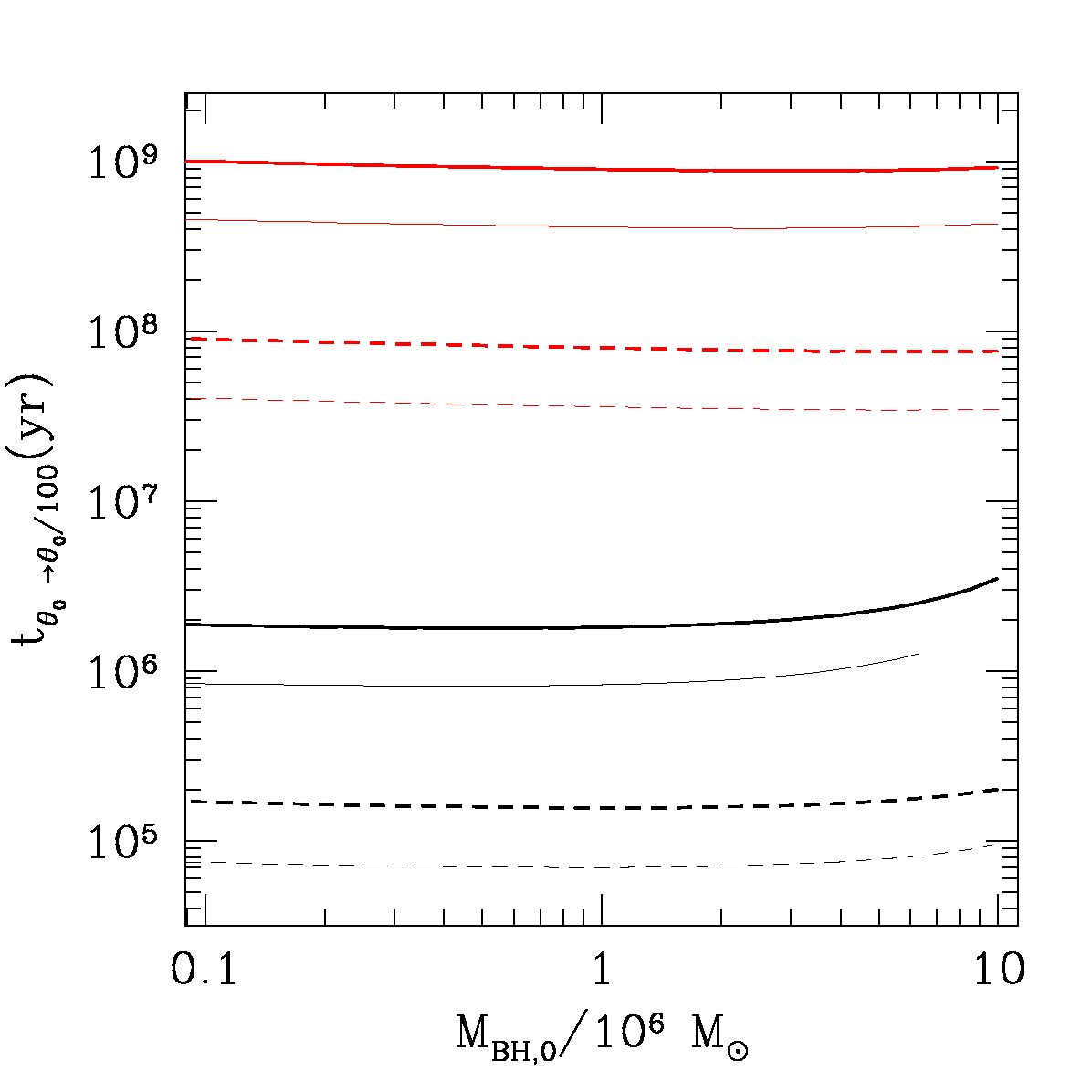}
    \caption{Alignment time (defined as the time needed for the relative 
inclination angle to reduce of two orders of magnitude, going from $\pi /6$ 
to $\pi /600$ ), as a function of the initial BH mass, $\mbho$, for
constant viscosity profile. 
The black lines refer to $f_\ed=1$ and the red ones to $f_\ed=0.001$; the solid lines are for initial spin parameter $a_0=0.9$ while the dashed ones $a_0=0.1$; finally, the thin lines represent $\alpha=0.09$ and the thick ones $\alpha=0.18$.}
    \label{fig:allt}
\end{minipage}
\hspace{10mm}
\begin{minipage}{0.45 \linewidth}
  \centering
     \includegraphics[width=82mm]{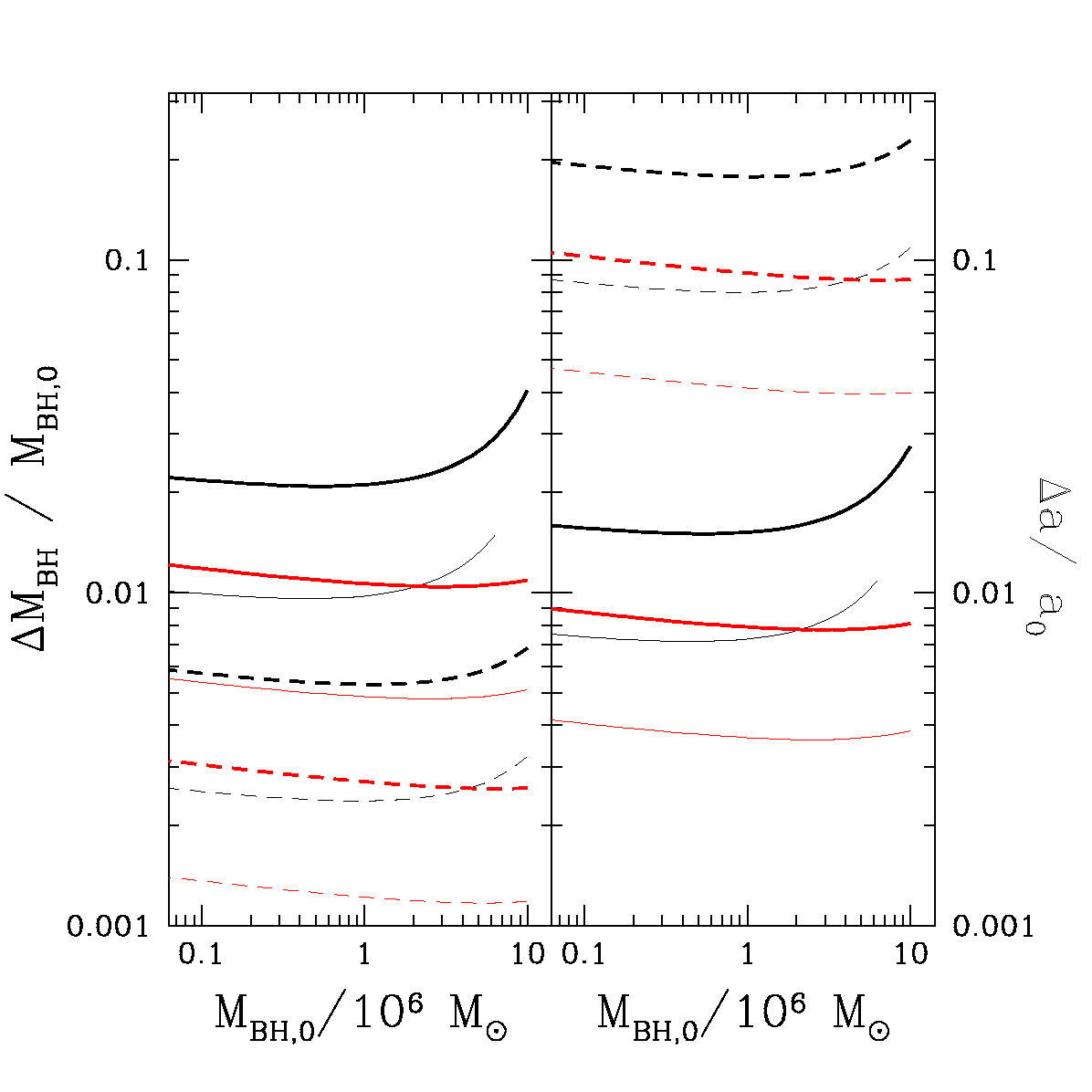}
    \caption{Mass and spin relative increase  during the alignment time 
(defined as the time needed for the relative inclination angle to reduce 
of two orders of magnitude, going from $\pi /6$ to $\pi /600$ ) as a 
function of the initial BH mass, for constant viscosity profile. 
Black (red) lines refer to $f_\ed=1$ 
($f_\ed=0.001$). Solid (dashed) lines are for $a_0=0.9$ ($a_0=0.1$);
finally, the thin (thick)  lines represent $\alpha=0.09$  ($\alpha=0.18$).}
    \label{fig:allms}
\end{minipage}
\end{figure*}

\begin{figure*}
  \includegraphics[width=144mm]{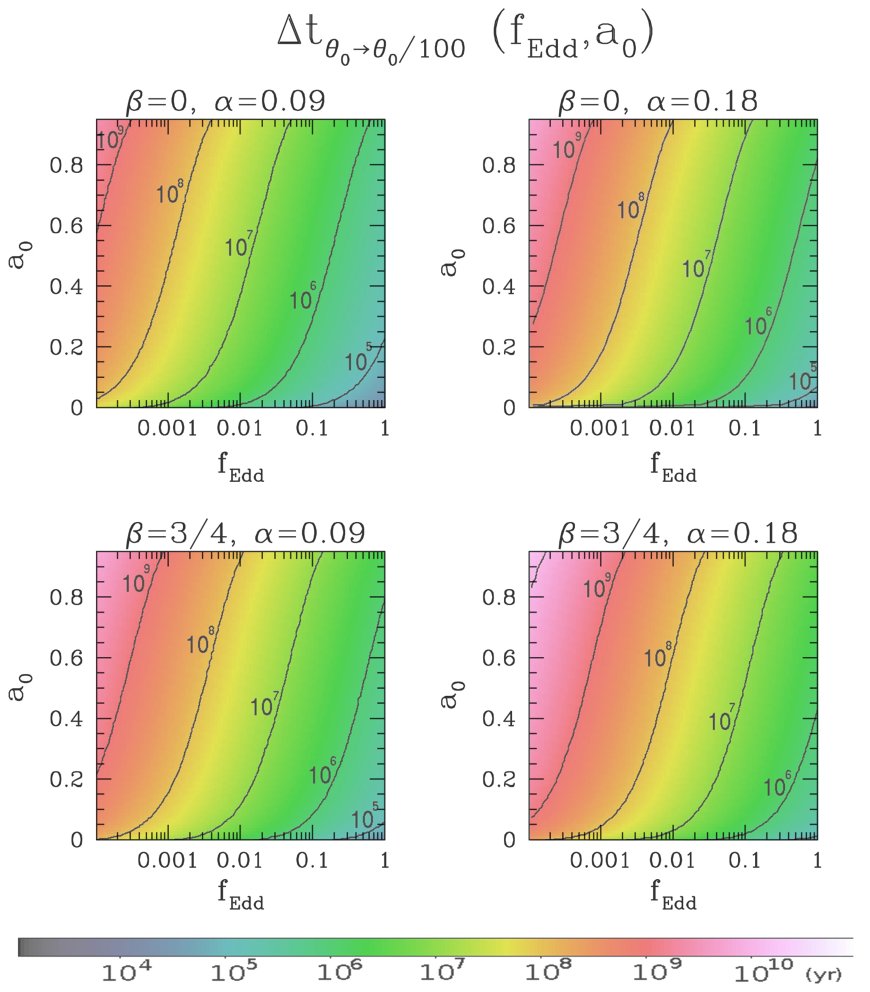}
  \caption{Color coded map of the aligment time (defined as the time necessary for the relative inclination angle to go from $\theta_\outo= \pi / 6$ to $\theta_\out= \pi / 600$) for a BH of $M_{\bh,0}=10^6 \msun$, as a function of the accretion rate expressed through the Eddinfton factor $f_\ed$ and of the initial BH spin parameter $a_0$. The colour scale represents $t_\al$ in years. Top (bottom) panels refer to the constant (power-law) viscosity profile. Left (right) panels refer $\alpha=0.09$ ($\alpha=0.18$). }
  \label{fig:tempi}
\end{figure*}
\begin{figure*}
  \includegraphics[width=144mm]{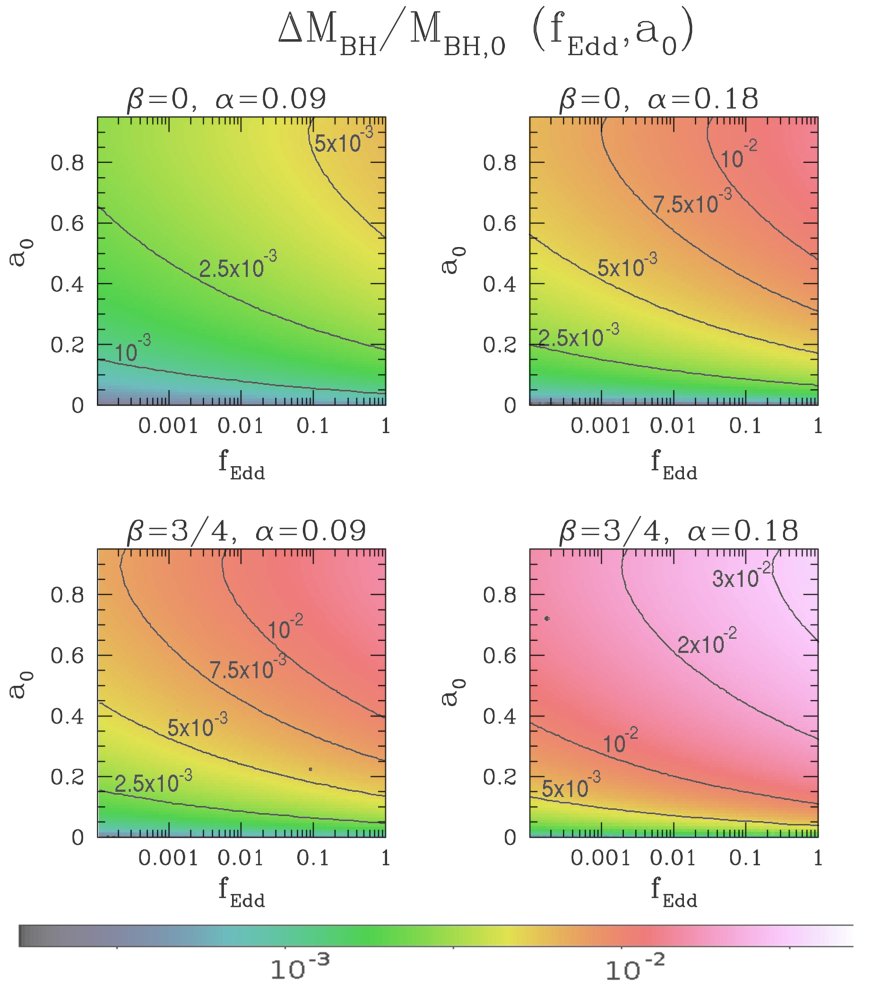}
  \caption{Color coded map of the relative increase of mass during the alignment time (defined as the time necessary for the relative inclination angle to go from $\theta_\outo= \pi / 6$ to $\theta_\out= \pi / 600$) 
for a BH of $M_{\bh,0}=10^6 \msun$, 
as a function of the accretion rate expressed through the Eddington factor $f_\ed$ and of the initial BH spin parameter $a_0$.
The colour scale represents $\Delta \mbh / \mbho$. Top (bottom) panels refer to the constant (power-law) viscosity profile. Left (right) panels refer $\alpha=0.09$ ($\alpha=0.18$).}
  \label{fig:masse}
\end{figure*}
\begin{figure*}
  \includegraphics[width=144mm]{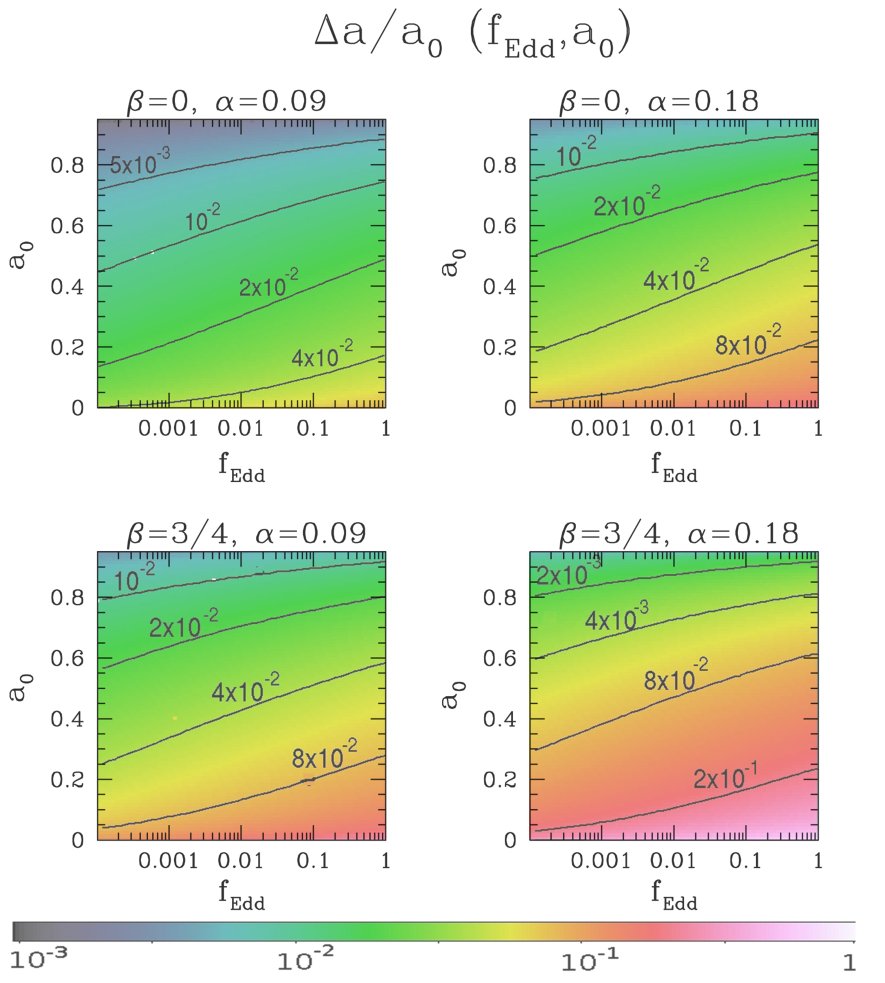}
  \caption{Color coded map of the relative increase of spin parameter during the alignment time (defined as the time necessary for the relative inclination angle to go from $\theta_\outo= \pi / 6$ to $\theta_\out= \pi / 600$) 
for a BH of $M_{\bh,0}=10^6 \msun$, 
as a function of the accretion rate expressed through the Eddinfton factor $f_\ed$ and of the initial BH spin parameter $a_0$. 
The colour scale represents $\Delta a/ a_0$. Top (bottom) panels refer to the constant (power-law) viscosity profile. Left (right) panels refer $\alpha=0.09$ ($\alpha=0.18$).}
  \label{fig:spin}
\end{figure*}

Here we explore more systematically how the fractional increases of $M_\bh$ 
and $a$, and the alignment time vary with initial mass $M_{\bh,0}$, 
spin ${a}_0$, $f_\ed$ and $\alpha$, for both constant and power-law viscosity 
profiles, fixing $\theta_{\rm out,0}= \pi/6$. 
The evolution is followed until $\theta_{\rm out}$ has decreased by a 
factor 100;
we define as $\Delta t_{\theta_0 \rightarrow \theta_0/100}$ 
the corresponding "alignment" time, computed self-consistently.
We also infer
from the numerical model the relative growths of BH mass 
$\Delta \mbh / \mbho$ and spin parameter $\Delta a / a_0$ 
during $\Delta t_{\theta_0 \rightarrow \theta_0/100}$.

Figure \ref{fig:allt} and Figure \ref{fig:allms} show the weak dependence 
of the alignment  time $\Delta t_{{\theta_0} \rightarrow \theta_0/100}$ 
on the initial BH mass $\mbho$ , and of the relative mass and
spin parameter increases, for eight different sets of the other parameters.  
Comparing numerical scaling factors for $M_6$ in $\Delta t_{\theta_0 
\rightarrow \theta_0/100}$ with that of expression 
(\ref{eqn:alignment timescale}), we notice again a good agreement, in 
particular for $f_\ed$ not too close to the Eddington limit and $\mbho \lesssim 10^6 \msun$.

By contrast, the alignment process is more sensitive on $f_\ed$, 
${a}_0$ and $\alpha$. 
Color-coded maps of $\Delta t_{{\theta_0} \rightarrow \theta_0/100}$ 
(Figure \ref{fig:tempi}), of $\Delta M/M_{\bh,0}$ (Figure \ref{fig:masse})
and $\Delta {a}/{a}_0$ (Figure \ref{fig:spin}) are constructed 
in the ${a}_0$ 
versus $f_\ed$ plane, varying the coefficient $\alpha$ and the viscosity law 
inside the accretion disc.\\
In Figure \ref{fig:tempi} we infer the
interval of the alignment time $\Delta t_{{\theta_0} \rightarrow \theta_0/100}$ (as inferred from the numerical model) of interest for the study of 
BH evolution. 
The alignment time  
can vary by many orders of magnitude from 
$\sim 10^5 {\rm yr}$ to $\sim 10^{10} {\rm yr}$, and it 
reveals strong dependencies both on the accretion rate and on the initial
spin parameter. 
In addition, smaller viscosities ($\alpha=0.09$) gives shorter timescales 
compared to higher viscosities ($\alpha=0.18$).
A simple comparison between alignment times 
$\Delta t_{{\theta_0}\rightarrow {\theta_0/100}}$ for different 
initial spin parameters, but identical $f_\ed$, reveals that the scaling 
factors for $a$ and $\eta_{0.1}(a)$ in equation (\ref{eqn:alignment timescale}) 
are in good agreement with numerical results.\\
Figure \ref{fig:masse} shows that the relative amount of mass 
accreted during the alignment process is small, compared 
to the initial BH mass. It varies between 
$\sim 10^{-3}$ and $\sim 10^{-2}$ for the constant viscosity profile, and
between $\sim 2.5 \times 10^{-3}$ and $\sim 3 \times 10^{-2}$ for the 
power-law viscosity profile.
Even if the accretion rate varies over four orders of magnitude, there are no
comparable variations for the relative BH mass growth, in fact 
a larger $f_\ed$ means a larger accretion rate, but it also reduces the alignment time.
The relative increase of the spin parameter $a$ is shown in Figure 
\ref{fig:spin}. The evolution of $a$ is the combination of different, and sometime opposite, tendencies: 
a highly spinning black hole requires a longer time to align, but the particles at its innermost stable orbit carry on the BH a smaller angular momentum. 
The spin modulus increases significantly  during the alignment for 
initially slowly rotating BHs and high accretion rates, typically with $ 5 \times 10^{-3} \lesssim \Delta a / a_0 \lesssim 8 \times 10^{-2}$ for a constant viscosity profile
and $ 10^{-2} \lesssim \Delta a / a_0 \lesssim 2 \times 10^{-1}$ for a
power-law profile.

\subsection{Counter-rotating case}

\begin{figure*}
  \centering
     \includegraphics[width=144mm]{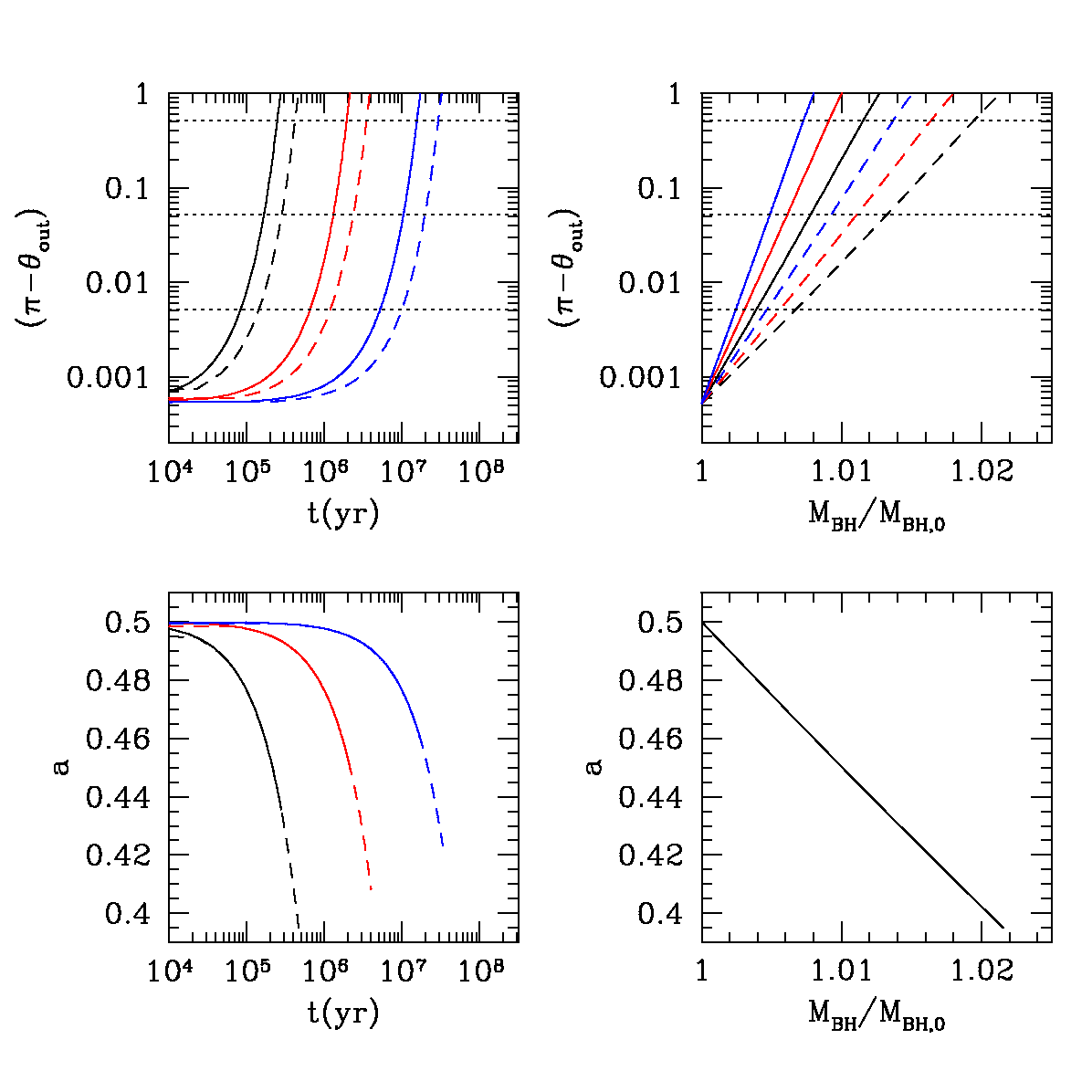}
    \caption{
Coupled evolution of the relative inclination angle $(\pi-\theta_\out)$, 
BH mass $\mbh$ and spin parameter $a$ for a counter-rotating disc. Solid 
(dashed) lines refer to constant (power-law) viscosity profile. Black lines 
to $f_\ed=1$, red lines to $f_\ed=0.1$, blue lines to $f_\ed=0.01$. Dotted 
horizontal lines which appear in top panels represent angles 
$\theta_\out / \theta_\outo=10,10^{2},10^{3}$. Initial configuration: 
$\mbho=10^6 \msun$, $a_0=0.5$, $f_\ed=0.1$ and $\alpha=0.09$, 
with $\pi-\theta_\outo= \pi / 600$.}
    \label{fig:vslcnt}
\end{figure*}

\begin{table*}
   \begin{center}
     \begin{tabular}{|c|c|c|c|c|c|}
       \hline
VP & $f_\ed$  & $(\pi - \theta_\out) / (\pi-\theta_\outo)$ & $\Delta t$   & $\Delta \mbh/M_{\bh,0}$ & $\Delta a/a_0$ \\
 & &  & ($10^6{\rm yr}$)&(in units of $10^{-2}$) & (in units of $10^{-2}$)\\
      \hline
      \hline
           &      & $10$   & 0.085  & 0.40  &  -3.99 \\
           &  1   & $10^2$ & 0.17   & 0.78  &  -7.81 \\
           &      & $10^3$ & 0.25   & 1.16  &  -11.5 \\
      \hline
           &      & $10$   & 0.66   & 0.31  & -3.12 \\
       C   &  0.1 & $10^2$ & 1.31   & 0.61  & -6.15 \\
           &      & $10^3$ & 1.96   & 0.91  & -9.11 \\
     
      \hline
           &      & $10$   & 5.29   & 0.25  & -2.49 \\
           & 0.01 & $10^2$ & 10.5   & 0.49  & -4.92 \\
           &      & $10^3$ & 15.7   & 0.73  & -7.31 \\
      \hline
      \hline
           &      & $10$   & 0.15   & 0.68  & -6.79 \\
           &  1   & $10^2$ & 0.29   & 1.33  & -13.1 \\
           &      & $10^3$ & 0.42   & 1.95  & -19.1 \\
       \hline
           &      & $10$   & 1.21   & 0.57  & -5.68 \\
       PL  & 0.1  & $10^2$ & 2.38   & 1.11  & -11.0 \\
           &      & $10^3$ & 3.52   & 1.64  & -16.2 \\     
      \hline
           &      & $10$   & 10.1   & 0.47  & -4.73 \\
           & 0.01 & $10^2$ & 19.9   & 0.93  & -9.24 \\
           &      & $10^3$ & 29.5   & 1.37  & -13.6 \\
      \hline
    \end{tabular}    
\caption{Summary of our parameters and results for the counter-rotating case; we consider viscosity coefficient $\alpha=0.09$ and initial inclination angle $\theta_\outo=\pi(1-1/6000)$,  both for constant (C) and power-law (PL) viscosity profiles (VP). 
The initial BH has $\mbho = 10^6 \msun$ and $a_{0} = 0.5$.
Accretion rate $f_\ed$ varies over three orders of magnitude and we record times needed to ($\pi-\theta_\out,0$) of a factor 10, 100 or 1000; we also report mass and spin relative variations.}
    \label{tab:angle variations cnt}
  \end{center}
\end{table*}

In this Section, we investigate the counter-rotating configuration for a
BH and its misaligned accretion disc, for initial values of $\theta_\outo$ 
close to $\pi$. 

As shown by \citet{ScheuerFeiler1996} and \citet{MartinPringleTout2007}, on 
the timescale $t_{\rm al}$ the BH spin again aligns with the outer regions 
of the accretion disc, if this disc is regularly and coherently fed.
Due to the Bardeen-Petterson effect, we expect the innermost part of the disc
(approximately within $R_\w$) to orbit in a plane which is perpendicular
to $\bfj_\bh$, with orbital angular momentum density ${\bf L}$ counter-aligned
with respect to the BH spin. In this BH-disc configuration, one of the major 
changes is in the radius of the innermost stable orbit, which increases due 
to the asymmetry seeded in the geodetic motion of particles in Kerr metrics. 
As a consequence,
the energy and the orbital angular momentum of particles at $R_{\rm ISO}$
increase, while the BH radiative efficiency decreases \citep[see, i.e.,][]{Wilkins1972,Bardeenetal1972}.
The Bardeen-Petterson timescale (\ref{eqn:tbp}) and 
the warp radius (\ref{eqn:Rw implicit}) have the same values as in the 
co-rotating case.
Since $t_\bp \ll t_\al$, the adiabatic approximation holds again, but 
the small deformation approximation\footnote{The functions $\psi$ and $\chi$, defined in equations (\ref{eqn:psi definition}) and (\ref{eqn:chi definition}), are invariant under the transformation $\theta_\out \rightarrow \left( \pi-\theta_\out \right)$. } has a limited validity, requiring $\theta_\out \sim \pi.$ 
In order to remain consistent
with the approximation scheme, we trace the alignment process, 
from $\pi$ to $\left( \pi - \pi/6 \right)$, only.

In the couter-rotating case and 
small deformation approximation (i.e. $\theta_\out \sim \pi$), the disc 
profile can be solved analytically.
We choose a reference frame $O''x''y''z''$ where $\hat{\bfj}_\bh=(0,0,-1)$ and
we solved equation (\ref{eqn:angular momentum}) for $\hat{\mathbf{l}}$ in it.
For constant viscosities, the function $W''(R/R_\w)={\hat l}''_{\rm x}+i{\hat l}''_{\rm y}$ describing the warp is 
\begin{equation} 
W''_{\rm C,cnt}=C \exp{\left(-\sqrt{2}~(1+i)\left( \frac{R}{R_\w}\right)^{-\frac{1}{2}} \right) }
\end{equation}
while for a power-law viscosity profile
\begin{equation}
\begin{aligned} 
& W''_{\rm PL,cnt}=D \left( \frac{R}{R_\w} \right)^{-\frac{1}{4}}\\
& \qquad \times \quad K_{{1}/{2(1+\beta)}}\left(\frac{\sqrt{2}(1+i)}{(1+\beta)} \left( \frac{R}{R_\w} \right)^{-\frac{1+\beta}{2}} \right). 
\end{aligned}
\end{equation}
We then apply the adiabatic approximation to study the coupled evolutions of
the system BH-disc.
The jointed evolutions of $\theta_{\rm out}$, $\mbh$ and $a$ 
are presented in Figure \ref{fig:vslcnt} and in Table 
\ref{tab:angle variations cnt}, for different accretion rates and viscosity 
profiles. The shorter timescales in the counter-rotating 
configuration stem from the the dependence of the alignment timescale on the spin modulus,
 $t_\al \propto {a}^{5/7}$.  Counter-rotating matter carries larger and opposite angular 
 momentum, reducing the spin modulus and the alignment timescale in the process.

\section{Discussion and conclusions}
\label{sec:conclusions}

In this paper, we followed the joint evolution of the mass $M_\bh$ 
and spin $\bfj_\bh$ of a BH inside a geometrically thin, extended accretion 
disc.
The BH spin is initially misaligned with the angular momentum of the disc 
in its outer regions.
On the short Bardeen-Petterson timescale, the disc responds  
to the Lense-Thirring precession, imposed by the BH spin, 
and propagates a warp that is maximum around $R_\w$; within this radius
matter orbits around the BH in a plane which is perpendicular to the BH spin.
According to angular momentum conservation, the warped disc interacts with
the BH spin and, on the longer alignment timescale, the BH aligns its spin 
to ${\hat \bfj}_{\rm disc,out}$.
In its outer regions, the disc is assumed to be fed by matter that flows
along a plane that keeps its coherence in direction ${\hat \bfj}_{\rm disc,out}$
for a sufficiently long timescale to allow for the gravitomagnetic
interaction to complete BH-disc alignment.
While doing so the BH is accreting matter and angular momentum from
the inner portion of the disc, which is aligned or anti-aligned to the BH spin.
Given the mismatch between the timescale for warp propagation and
the alignment time \citep{ScheuerFeiler1996, NatarajanPringle1998}, 
we devised a method that enabled us to
follow, in the small-deformation approximation, the co-evolution of the BH mass 
and spin in a self-consistent manner, carrying out a large survey of the 
parameter space and a critical review of the used approximations.

It is found that, considering an initial small relative inclination angle
($\theta_\outo \lesssim \pi/6$, small deformation approximation), matter in the 
inner part of the accretion disc has orbital angular momentum density 
parallel to $\bfj_\bh$. 
The gravitomagnetic interaction of the BH with this warped accretion disc and 
their coupled evolution bring the BH into alignment with the outer 
regions of the disc, i.e. $\theta_\out(t) \rightarrow 0$.
The timescale $t_\al$ of equation (\ref{eqn:alignment timescale}) gives a good 
estimate of the BH-disc alignment time 
for an $e$-folding reduction of the angle
of misalignment, in very good agreement with numerical results.
For a maximally rotating Kerr BH accreting at the Eddington rate, 
$t_\al\sim 10^{5-6}$ yr, depending on the viscosity
parameter $\alpha$ and on the viscosity profile model, in agreement with 
early findings by \citet{NatarajanPringle1998}. 
On the other hand, environments where the accretion rate is extremely low 
imply longer alignment timescales, 
as $t_\al \propto {\dot M}^{-32/35}$.
In the explored BH mass range, the alignment time displays a weak dependence 
on $\mbho$: fixed all the other parameters, alignment of a 
$10^7\,\msun$ BH occurs, on average, at the same pace of a $10^5\msun$ BH.
The BH mass and spin modulus increase during alignment, but their fractional 
increases are modest. After surveying a wide parameters space, we find that 
$0.1 \% \lesssim \Delta M_\bh/M_{\bh,0}\lesssim 3\%$ 
while the spin parameter increases by 
$ 0.5 \% \lesssim \Delta {a}/{a}_0 \lesssim 20 \% $.

Starting with an almost anti-parallel BH-disc configuration
($\theta_\outo \approx \pi$), the orbital angular momentum density of the inner
part of the disc is initially counter-aligned with respect to the BH spin.
Nevertheless, the BH still tends to reduce the degree of misalignment 
(i.e. $\theta_\out(t)$ decreases), because of the nature 
of the gravitomagnetic interaction \citep[see also][]{ScheuerFeiler1996, 
MartinPringleTout2007}.
The accretion of matter with opposite angular momentum at $R_{\rm ISO}$
decreases $\bfj_\bh$ and $a$ with higher rates, compared with 
their growths in the specular co-rotating case. 
Since $t_{\rm al} \propto a^{5/7}$,
the alignment process is then more efficient, i.e. the angle reduction speed
is higher. 
Comparing decreases of the relative inclination angle 
$\theta_\out$ symmetric with respect to $\pi/2$, we find that
the fractional decrease of the spin parameter in the counter rotating case 
is, in modulus, higher than in the specular co-rotating case while the mass 
relative increase is slightly lower.
BH spin flip, due to $\theta_{\rm out}$ reduction below $\pi/2$, will 
occur in this extended disc
when $a$ will reach its minimum value.
At that time the jet of relativistic particles (if present) will
cross the warped disc, likely affecting the subsequent BH-disc evolution and
the BH feeding. This process deserves a separate investigation. 

It is still poorly known whether a spinning BH in an active galactic nucleus is
fed through a disc that maintains its angular momentum direction 
{\it stable} over a Salpeter timescale $t_{\rm S}$.
Two opposite, still plausible scenarios, have been proposed and discussed.
\citet{NatarajanPringle1998} speculated that 
the stability of jets in radio-loud AGNs
requires a long-lived phase of stable accretion capable to  
maintain spatial coherence, i.e. a fixed direction of $\bfj_{\rm disc, out},$ 
for a time as long as $10^8$ yr.
By contrast, \citet{KingPringle2006, KingPringle2007, KingPringleHofmann2008} 
speculated recently that AGN activity, triggered by gas-rich major
mergers, is chaotic in nature even within a single merger event,  
i.e. is occurring through a sequence of uncorrelated 
short-lived accretion episodes. In their picture the 
corresponding discs, truncated by their
own-self gravity, continuously change their inclination and feed the BH
on their consumption timescale. 
Under these circumstances  
the BH spin modulus is seen to either increase or decrease at random 
clustering around small average values ${a}\sim 0.1-0.3$. 
This model would simultaneously explain the relatively low radiative efficiency
of the quasar population as 
inferred from the background light \citep[e.g.,][]{Merloni2004,MerloniHeinz2008}, 
and the possibility of growing BH as massive as $10^9 \msun$ from small BH 
seeds already at redshift $z\sim 6$ \citep{KingPringle2006}.

Isolated discs, truncated by their own self-gravity, carry 
a well defined disc angular momentum and are accreted by the
BH on a finite timescale.
Starting with a misaligned BH-disc configuration, the BH spin changes direction
significantly only if 
(i) the alignment time is shorter than the disc consumption time, $t_\al<t_{\rm disc}$;
and (ii)  the magnitude of the disc angular momentum is comparable to the BH spin magnitude, i.e.  $J_{\rm disc} \gtrsim J_{\rm BH}$. 
The first condition is verified for the whole parameter range explored in this paper. 
The estimate $J_{\rm disc} \gtrsim J_{\rm BH}$ depends instead sensitively upon $R_\out$.
Equation \ref{eqn:ratio between momenta} establishes that isolated discs around large BHs 
truncate at $R_\out$ such that $J_{\rm disc} < J_{\rm BH}$.  Condition (ii) is satisfied for BH masses $ \lesssim 3 \times 10^7 \msun$.
We note here that since we model our discs using the Shakura-Sunyaev solution
for Kramer's opacity, we cannot rigourously estimate $R_\out$, and $J_{\rm disc}$,
for BHs with mass $\lesssim10^5-10^6 \msun$. An extension of disc solutions to 
different, self-consistent opacities is non trivial 
\citep{HureetalI1994,HureetalII1994} and we postpone a detailed analysis to 
future work. 

BHs with masses $\lesssim 3\times10^7$ align efficiently in discs truncated by their own self-gravity, implying  alignment also
in the case of stochastically fed AGNs\footnote{Here we are not considering 
accretion events involving a disc with a very small amount of
mass, i.e. below $\sim (H/R)M_\bh$. This light accretion disc has an outer radius much
smaller than (\ref{eqn:r out}) and thus carries an angular momentum $J_{\rm disc}\ll J_\bh$.
As a consequence, the disc has a very short consumption timescale, and the alignment process is active
for a very short period of time. Therefore the alignment of $\bfj_\bh$ 
around $\bfj_{\rm tot}$ (close to $\bfj_\bh$) is expected to be unimportant.}.
where not only  $a$ fluctuates with time, but also the direction of 
the BH spin continually changes due to the rapidity of the alignment process.
By contrast, rapidly spinning ($a\sim 1$) heavier BHs with 
$M_{\rm BH} \gtrsim 10^8 \msun$ have truncated discs that 
carry little angular momentum compared with $J_\bh$.
In this case alignment is uneffective and the orientation of the BH spin is
not influenced significantly by the surrounding short-lived disc. 

In light of these findings, the vector $\bfj_\bh$ appears to carry 
precious information on the orientation of
the plane through which the BH has been fed, and on 
whether accretion has been long-lived and coherent or short-lived and random. 

The method developed in the paper is sufficiently versatile that it will 
be implemented in numerical simulations
describing  the process of pairing of dual BHs in circumnuclear discs during their on-fly accretion 
(Dotti et al., in preparation) to improve upon the speculation \citep{Bogdanovicetal2007} that, in gas-rich galaxy mergers,
binary BHs have time to align their spin orthogonally to their orbital plane, as discussed in
\cite{Escalaetal2005,Dottietal2006,Mayeretal2007,Dottietal2007,Dottietal2009,ColpiDotti2009}. 
The spin-orbit configuration is relevant to study the impact of BH recoils,  that occur 
after two BHs have coalesced \citep[see, e.g.,][]{Pretorius2007}.
Detection of gravitational waves, emetted by coalescing BHs, with the 
{\it Laser Interferometer Space Antenna} ({\it LISA})
\citep{Benderetal1994, HilsBender1995} will be able to 
constrain the moduli and the directions of the coalescing BHs spins
\citep{Vecchio2004,LangHughes2006}.

\section{ACKNOWLEDGMENTS}
We wish to thank Vittorio Gorini, Sergio Cacciatori, Alberto Sesana, 
Bernadetta Devecchi, Oliver Piattella and Luca Rizzi for usefull discussions 
and suggestions.

\bibliographystyle{mn2e}

\bibliography{bibalignment}

\appendix

\section[]{Explicit expressions of the gravitomagnetic torque}

{\bf  Constant viscosity}: For the constant viscosity model, the disc profile is described by (\ref{eqn:solution for W, nu constant}) and the equation (\ref{eqn:torque integral}) becomes 
\begin{equation} \label{eqn:gravitomagnetic torque: constant viscosities}
\begin{aligned}
& \delta(J_{\bh,x}+iJ_{\bh,y})_{\rm gm}= \\
& \qquad (1-i)\frac{2 \sqrt{2} A}{3}\sqrt{G\mbh} \frac{G \dot{M}J_\bh}{A_{\nu_1}c^2 R_\w^{5/4}}t_\bp(R_\w) \\
& \qquad \times \left. \exp{\left(-\sqrt{2}(1-i)\left(\frac{R}{R_\w} \right)^{-1/2}  \right)} \right|^{R_\out}_{R_\ISO}.
\end{aligned}
\end{equation}
{\bf Power-law viscosity}: 
For the power-law case (\ref{eqn:Sa-Su viscosity 1}) with exponent $\beta$, the disc profile is given by $W_{\rm PL}$, defined as (\ref{eqn:solution for W, nu power-law}). In this case, equation (\ref{eqn:torque integral}) was integrated by \citet{MartinPringleTout2007}:
\begin{equation} \label{eqn:gravitomagnetic torque: power-law viscosities}
\begin{aligned}
& \delta(J_{\bh,x}+iJ_{\bh,y})_{\rm gm}= -i~\frac{8G \dot{M} J_\bh\sqrt{G\mbh}}{3(1+\beta)A_{\nu_1}c^2} \\
& \qquad \times B~\left(\frac{\sqrt{2}}{1+\beta}(1-i) \right)^{-\frac{4\beta+3}{2(1+\beta)}} R_\w^{-\left( \beta+\frac{1}{2}\right)}\\
&  \qquad \times t_\bp(R_\w)~\int_{z_{\rm in}}^{z_\out}z^{\frac{2\beta+1}{2(1+\beta)}}K_{\frac{1}{2(1+\beta)}}(z)~dz
\end{aligned}
\end{equation}
where $z$ is a new complex variable, defined as
\begin{equation}
z=\frac{\sqrt{2}}{1+\beta}(1-i)\left(\frac{R}{R_\w} \right)^{-\frac{1+\beta}{2}}.
\end{equation}
Assuming that
\begin{equation} \label{eqn:integral maprito}
\begin{aligned}
& \int_{z_{\rm in}}^{z_\out}z^{\frac{2\beta+1}{2(1+\beta)}}K_{\frac{1}{2(1+\beta)}}(z)~dz \approx \\
& \qquad \int_0^{(1-i)\infty}z^{\frac{2\beta+1}{2(1+\beta)}}K_{\frac{1}{2(1+\beta)}}(z)~dz = 2^{-\frac{1}{2(1+\beta)}} \\
& \qquad \times \Gamma\left( \frac{1+2\beta}{2(1+\beta)} \right)
\end{aligned}
\end{equation}
we can rewrite the infinitesimal gravitomagnetic spin variation as
\begin{equation} \label{eqn:explicit torque calculation PL}
\frac{\delta(J_{\bh,x}+iJ_{\bh,y})_{\rm gm}}{J_BH}= i^{-\frac{1}{4(1+\beta)}}~\frac{t_\bp(R_\w)}{T_{\rm PL}}
\end{equation}
where 
\begin{equation}
\begin{aligned}
& T_{\rm PL}^{-1}= \frac{4G \dot{M}\sqrt{G\mbh}}{3~A_{\nu_1}c^2}~B~~\left(\frac{\sqrt{2}}{1+\beta}\right)^{-\frac{2\beta+1}{2(1+\beta)}} \\
& \qquad \times R_\w^{-\left( \beta+\frac{1}{2}\right)} 2^{-\frac{2\beta +3}{4(1+\beta)}}~\Gamma\left( \frac{1+2\beta}{2(1+\beta)} \right).
\end{aligned}
\end{equation}
\citet{MartinPringleTout2007} estimate the alignment timescale as
\begin{equation} 
\label{eqn:Martin scale}
t_{\al,M} = \frac{T_{\rm PL}}{\cos \left( \frac{\pi}{4(1 + \beta)}  \right)}.
\ee 

\label{lastpage}

\end{document}